\documentclass[english]{article}
\usepackage[T1]{fontenc}
\usepackage[latin9]{inputenc}
\usepackage{geometry}
\usepackage{amsmath}
\usepackage{graphicx}
\usepackage{setspace}
\usepackage{float}
\usepackage{babel}
\begin{document}
\begin{singlespace}

\title{Local Thermometry of Neutral Modes on the Quantum Hall Edge}

\author{\noindent Vivek Venkatachalam$^{\dagger1}$, Sean Hart$^{\dagger1}$,
Loren Pfeiffer$^{2}$, Ken West$^{2}$, Amir Yacoby$^{1}$}

\date{\noindent $^{1}$Department of Physics, Harvard University, Cambridge,
MA, USA \\
$^{2}$ Department of Electrical Engineering, Princeton University,
Princeton, NJ, USA\\
$^{\dagger}$These authors contributed equally to this work}

\maketitle
\textbf{Summary: }Quantum dots, used as local thermometers, detect
upstream heat transport in a $\nu=2/3$ fractional quantum Hall edge
state, even when the state is carrying no net charge.
\begin{abstract}
A system of electrons in two dimensions and strong magnetic fields
can be tuned to create a gapped 2D system with one dimensional channels
along the edge. Interactions among these edge modes can lead to independent
transport of charge and heat, even in opposite directions. Measuring
the chirality and transport properties of these charge and heat modes
can reveal otherwise hidden structure in the edge. Here, we heat the
outer edge of such a quantum Hall system using a quantum point contact.
By placing quantum dots upstream and downstream along the edge of
the heater, we can measure both the chemical potential and temperature
of that edge to study charge and heat transport, respectively. We
find that charge is transported exclusively downstream, but heat can
be transported upstream when the edge has additional structure related
to fractional quantum Hall physics. 

\end{abstract}
When a two-dimensional electron system (2DES) is subject to a strong
perpendicular magnetic field and tuned such that the ratio of electrons
to magnetic flux quanta in the system ($\nu$) is near certain integer
or fractional values, the bulk of the system develops a gap due to
either quantization of kinetic energy (the integer quantum Hall, or
IQH, effect) or strong correlations arising from non-perturbative
Coulomb interactions (the fractional quantum Hall, or FQH, effect)
\cite{Girvin1987}. While the bulk (2D) is gapped and incompressible,
the edge (1D) of the system contains compressible regions with gapless
excitations that carry charge chirally around the system, in a direction
determined by the external magnetic field. Compressible edge states
have gained more attention recently due to their ability to serve
as a bus for quasiparticles that exist in exotic FQH phases\cite{Stern2006,Bonderson2006}.
These edges, however, can have considerable internal structure that
is not apparent from bulk transport measurements.

The spatial structure of edges is dictated by the interplay between
the external confining potential which defines the edge, an additional
harmonic confinement from the magnetic field, and Coulomb interactions.
It was predicted \cite{Chklovskii1992} and verified \cite{Zhitenev1993,Hwang1993,Yacoby1999}
that for a smooth, topgate-defined confining potential, it is energetically
favorable for the electron density to redistribute slightly to create
alternating compressible and incompressible strips. This has the effect
of spatially separating edges corresponding to different filling factors.
Such an effect is not present in sharper edges \cite{Huber2005}.

Perhaps more surprising than this spatial structure is the possibility
of modes that carry energy (or heat) upstream, even as the magnetic
field carries the injected charge downstream. The edge of the $\nu=2/3$
FQH state was originally predicted to consist of a $\nu=1$ edge of
electrons going downstream with a $\nu=1/3$ edge of holes going upstream
\cite{MacDonald1990,Wen1990}. However, this edge structure would
suggest a two-terminal conductance of $G_{2T}=\frac{4}{3}\frac{e^{2}}{h}$.
Scattering between the edges would lead to non-universal values in
the range of $\frac{2}{3}\frac{e^{2}}{h}\leq G_{2T}\leq\frac{4}{3}\frac{e^{2}}{h}$.
Experimentally, however, no such two terminal conductance has been
measured. Direct approaches to look for upstream charge transport
in the time domain have similarly turned up no evidence \cite{Ashoori1992}.
This motivated a picture in which disorder induces scattering and
equilibration between the edges, forcing the charge to travel exclusively
downstream. Heat, however, would be allowed to travel diffusively
upstream and downstream, leading to a nonzero thermal Hall conductivity
and partial upstream heat transport at $\nu=2/3$ \cite{Kane1994,Kane1997}. 

Evidence for upstream heat transport in a $\nu=2/3$ edge was recently
obtained by performing modified shot noise measurements \cite{Bid2010}.
Our approach studies the same state by directly placing thermometers
upstream and downstream of a current-source heater to observe charge
and heat transport along the edge.

As our heater, we use a lithographically fabricated quantum point
contact (QPC), tuned to the tunneling regime (Fig. 1C). Tunneling
of electrons through this QPC at elevated energy locally excites the
outermost compressible component of a gate-defined edge. We then place
quantum dots 20 $\mu\textrm{m}$ upstream and downstream of the QPC
to measure charge and heat transport (Fig. 1A). The edge itself is
defined by a separate pair of gates (green in Fig. 1A), and the perpendicular
magnetic field defines a clockwise charge-propagation direction (with
respect to Figure 1). All measurements were carried out in a dilution
refrigerator with a minimum electron temperature of $20$ mK, measured
with Coulomb blockade thermometry. 

To first characterize the structure of the edge that we are tunneling
charge into, we energize a subset of gates upstream (blue) and downstream
(red) of the central QPC to create additional point contacts that
serve as imperfect voltage probes ($R\sim100\ \textrm{k}\Omega$).
This ensures that we only measure the chemical potential of the outermost
edge component \cite{Wees1989}. Current is injected through the central
QPC ($10$ pA sourced through O3 and drained at O6). The upstream
chemical potential, $V1-V7$, was observed to be immeasurably small
in all measurements, indicating that no charge is transported upstream
on a 20 $\mu\textrm{m}$ scale. The downstream chemical potential,
$V5-V7$, can be used to determine the resistance of the edge connecting
the source to the probe (the {}``local Hall'' resistance $R_{L}$).
This resistance is plotted in blue in Figure 2. Additional measurement
details can be found in the online supplement.

For magnetic fields ($B$) between 2 T and 8 T, the measured value
$R_{L}=1\frac{h}{e^{2}}$ indicates that the charge is carried between
the injector and detector by electronic modes that behave similarly
to an IQH $\nu=1$ edge. Inner edges can (and must, at fields below
6 T) be present, as can be seen by comparing $R_{xy}$ with $R_{L}$.
These inner edges, however, do not carry any of the injected charge.
Above 8 T, we find that $R_{L}$ is quantized to $R_{L}=\frac{2}{3}\frac{h}{e^{2}}$
even though the bulk is at $\nu=1$\cite{Kouwenhoven1990}. This suggests
that the edge has additional structure consisting of alternating compressible
and incompressible regions which are spatially separated, as indicated
in Figure 2 (IV). In this situation, we only access the outermost
edge of the incompressible $\nu=2/3$ strip located outside the $\nu=1$
bulk. The robust quantization that we observe indicates that no charge
leaks out of this outermost $\nu=2/3$ edge over the 20 $\mu\textrm{m}$
separating the injector from the detector.

The edge-deflecting gates (green in Figure 1A) can be deenergized
to deflect the edges into floating ohmic contacts located 250 $\mu\textrm{m}$
away (O2 and O4), where they will chemically equilibrate and thermally
cool (though some equilibration and cooling may occur before the edges
reach the ohmic contacts). If we repeat this charge transport measurement
with the deflector gates deenergized, we continue to monitor no upstream
charge transport. However, the downstream resistance is observed to
match exactly the bulk value of $R_{xy}$, plotted in black in Figure
2. This indicates that our deflection process does, indeed, force
all edges to fully chemically equilibrate in ohmic contacts O2 and
O4, providing an important control for the heat transport measurements
discussed below. 

To characterize heat transport, we energize all of the gates upstream
and downstream of the central QPC to form quantum dots, which serve
as thermometers to measure the temperature of the edge. This is similar
to another recent spectroscopic approach \cite{Altimiras2009,Altimiras2010,Takei2010}.
The width of the Coulomb blockade peak as a function of gate voltage
can be translated into the temperature of the leads (Fig. 1B, details
in SOM A).

With the thermometers active, we inject current through the QPC set
to an average transmission of 15\% to create a non-equilibrium population
in the outermost edge (Fig. 1C). The low transmission ensures that
we inject solely electrons into the edge (no FQH edges are fully transmitted).
These energetic electrons, however, are not necessarily the elementary
excitations of the edge and will therefore excite the natural edge
modes as they decompose. By increasing the bias across the QPC, we
vary the current (and therefore the power) being delivered to the
edge. We monitor both the chemical potential and temperature of the
edge at the upstream and downstream dots (Fig. 1B and 1D). 

Measurements are first performed with the deflector gates energized,
to measure heat transport associated with the edge (red and blue curves
in rows 1 and 2 of Fig. 3). We then repeat the procedure with the
deflector gates off, to measure any background heating not associated
with the edge (cyan and magenta curves in rows 1 and 2 of Fig. 3).
The difference between these two temperatures gives us a measure of
the excess heat carried by the edge (bottom row in Fig. 3, red is
downstream and blue is upstream temperature).

At the two lowest fields that were measured (2.41 T and 3.8 T), our
charge transport measurements indicate that we are injecting charge
into a $\nu=1$ edge sitting outside an incompressible bulk at filling
$\nu=3$ or $\nu=2$ respectively. This is depicted schematically
in Fig. 2 (I,II) and in Fig. 4 (II). By monitoring the chemical potential
as we vary the injected power, we find that charge is carried exclusively
by the outermost $\nu=1$ edge over the entire range of measurement
(SOM A).

At 2.41 T, when the bulk is at $\nu=3$, there is no measurable background
heating either upstream or downstream. When the deflectors are turned
on, we find heating downstream but none upstream. When the bulk is
at $\nu=2$, we find about 2-3 mK of background heating that is perfectly
cancelled in the upstream direction. Thus, in both cases, we find
that heat carried by edge modes is transported exclusively downstream.
While this strict downstream heat transport in the IQH regime is expected
and matches previous measurements \cite{Granger2009,Bid2010}, surprisingly,
the magnitude of the temperature observed does not agree with what
one would expect from quantized thermal transport (assuming an equilibrated
edge):
\[
K_{H}\equiv\frac{\partial J_{E}}{\partial T}=n\frac{\pi^{2}}{3}\frac{k_{B}^{2}}{h}T\implies T=\frac{\sqrt{6hJ_{E}/n}}{\pi k_{B}},
\]
where $J_{E}$ is the power carried by the edge and $n$ is the number
of IQH edges participating in transport \cite{Kane1997}. At $\nu=2$,
for an injected power of 350 fW, we expect an edge temperature between
430 mK and 608 mK, depending on how well the two edges thermally equilibrate
($n=2$ or $n=1$). Our measured temperature of 30 mK indicates that
a substantial quantity of heat is transferred out of the edge\cite{Altimiras2010}.
We can model the behavior of heat transport for out-of-equilibrium
Fermi systems (SOM B), which indicates a similar temperature deficiency.
Both models, however, give the correct shape for the temperature versus
power curves presented in Figure 3.

At the highest measured field, 8.3 T, charge transport (Fig. 2) indicates
that we have an incompressible $\nu=2/3$ strip outside a $\nu=1$
bulk, depicted schematically in Fig. 2 (IV). Here we see substantially
more background heating, both upstream and downstream, but after subtracting
contributions from the bulk (deflectors energized) we still find an
upstream temperature rise of 5 mK at 300 fW, compared to a downstream
rise of 11 mK. Such upstream heating is consistent with the predicted
diffusive conductivity of the outer $\nu=2/3$ edge\cite{Kane1997},
though the assymmetry between upstream and downstream temperatures
suggests that the inner $\nu=1\rightarrow2/3$ edge (which carries
heat preferentially downstream) is partially participating in heat
transport.

At the second highest measured field, 6.2 T, one would expect, based
on charge transport, behavior similar to what we find when the bulk
is at $\nu=2$ or $\nu=3$, with all heat being carried downstream
by the integer $\nu=1$ edge. Instead, we find behavior similar to
what was observed at 8.3 T, with heating both upstream and downstream
and a slight asymmetry between the two. This surprising result can
be understood if we allow for the presence additional structure in
the $\nu=1$ edge that does not affect charge transport. Perhaps the
simplest such structure would be the presence of an incompressible
strip of $\nu=2/3$, much like what we see at 8.3 T, but with charge
equilibrating between the two separated edges of this strip (Fig.
2 (III)). With these edges equilibrated, we measure a local Hall resistance
of $R_{L}=\frac{h}{e^{2}}$. However, the diffusive heat transport
provided by the outer $\nu=2/3$ edge could still carry heat to the
upstream thermometer (edge IV in Fig. 4). Additional evidence for
such an edge structure is presented in SOM D. Importantly, this mechanism
of upstream heating by an apparent $\nu=1$ edge would not be universal
and would depend sensitively on the spatial reconstruction of that
edge. A sharper mesa-defined edge with a larger density gradient \cite{Granger2009,Bid2010}
or a lower-mobility 2DES may not allow an incompressible strip of
$\nu=2/3$ to form outside the $\nu=1$ bulk. In the online supplement
(SOM D), we present a device with a mesa-defined edge that shows no
upstream heat transport at 6.2 T (edge III in Fig. 4).

By studying the charge and heat transport properties of the outermost
component of a gate-defined quantum Hall edge, these measurements
paint a picture in which such edges contain considerable structure.
Charge transport along the edge shows that correlated FQH modes can
exist outside an IQH bulk. Even when these charge signatures are not
present (Fig. 2 (III) and edge IV in Fig. 4), heat transport suggests
that density reconstructions can still create additional edge components
that carry heat upstream. In addition to this, by separating bulk
and edge contributions, we have been able to observe bulk heat transport
at $\nu=1$ which is absent at $\nu=2$ and $\nu=3$, the origin of
which remains an open question.

More generally, our system provides a framework to extract quantitative
information about charge and heat transport at the boundary of any
two-dimensional topological insulator. Such a system can be essential
to discriminate between topological states of matter that have identical
charge transport behavior. For example, with the $\nu=5/2$ FQH state,
the presence or absence of these neutral modes would allow us to discriminate
between distinct ground states that are particle-hole conjugates of
each other \cite{Lee2007,Levin2007}. 

\bibliographystyle{Science}
\bibliography{bibliography}

\begin{thebibliography}{10}

\bibitem{Girvin1987}
S.~Girvin, R.~Prange, {\it The Quantum Hall Effect\/} (Springer, 1987).

\bibitem{Stern2006}
A.~Stern, B.~I. Halperin, {\it Phys. Rev. Lett.\/} {\bf 96}, 016802 (2006).

\bibitem{Bonderson2006}
P.~Bonderson, A.~Kitaev, K.~Shtengel, {\it Phys. Rev. Lett.\/} {\bf 96}, 016803
  (2006).

\bibitem{Chklovskii1992}
D.~B. Chklovskii, B.~I. Shklovskii, L.~I. Glazman, {\it Phys. Rev. B\/} {\bf
  46}, 4026 (1992).

\bibitem{Zhitenev1993}
N.~B. Zhitenev, R.~J. Haug, K.~v. Klitzing, K.~Eberl, {\it Phys. Rev. Lett.\/}
  {\bf 71}, 2292 (1993).

\bibitem{Hwang1993}
S.~W. Hwang, D.~C. Tsui, M.~Shayegan, {\it Phys. Rev. B\/} {\bf 48}, 8161
  (1993).

\bibitem{Yacoby1999}
A.~Yacoby, H.~Hess, T.~Fulton, L.~Pfeiffer, K.~West, {\it Solid State
  Communications\/} {\bf 111}, 1 (1999).

\bibitem{Huber2005}
M.~Huber, {\it et~al.\/}, {\it Phys. Rev. Lett.\/} {\bf 94}, 016805 (2005).

\bibitem{MacDonald1990}
A.~H. MacDonald, {\it Phys. Rev. Lett.\/} {\bf 64}, 220 (1990).

\bibitem{Wen1990}
X.~G. Wen, {\it Phys. Rev. Lett.\/} {\bf 64}, 2206 (1990).

\bibitem{Ashoori1992}
R.~C. Ashoori, H.~L. Stormer, L.~N. Pfeiffer, K.~W. Baldwin, K.~West, {\it
  Phys. Rev. B\/} {\bf 45}, 3894 (1992).

\bibitem{Kane1994}
C.~L. Kane, M.~P.~A. Fisher, J.~Polchinski, {\it Phys. Rev. Lett.\/} {\bf 72},
  4129 (1994).

\bibitem{Kane1997}
C.~L. Kane, M.~P.~A. Fisher, {\it Phys. Rev. B\/} {\bf 55}, 15832 (1997).

\bibitem{Bid2010}
A.~Bid, {\it et~al.\/}, {\it Nature\/} {\bf 466}, 585 (2010).

\bibitem{Wees1989}
B.~J. van Wees, {\it et~al.\/}, {\it Phys. Rev. Lett.\/} {\bf 62}, 1181 (1989).

\bibitem{Kouwenhoven1990}
L.~P. Kouwenhoven, {\it et~al.\/}, {\it Phys. Rev. Lett.\/} {\bf 64}, 685
  (1990).

\bibitem{Altimiras2009}
C.~Altimiras, {\it et~al.\/}, {\it Nature Physics\/} {\bf 6}, 34 (2009).

\bibitem{Altimiras2010}
C.~Altimiras, {\it et~al.\/}, {\it Phys. Rev. Lett.\/} {\bf 105}, 226804
  (2010).

\bibitem{Takei2010}
S.~Takei, M.~Milletar\`\i, B.~Rosenow, {\it Phys. Rev. B\/} {\bf 82}, 041306
  (2010).

\bibitem{Granger2009}
G.~Granger, J.~P. Eisenstein, J.~L. Reno, {\it Phys. Rev. Lett.\/} {\bf 102},
  086803 (2009).

\bibitem{Lee2007}
S.-S. Lee, S.~Ryu, C.~Nayak, M.~P.~A. Fisher, {\it Phys. Rev. Lett.\/} {\bf
  99}, 236807 (2007).

\bibitem{Levin2007}
M.~Levin, B.~I. Halperin, B.~Rosenow, {\it Phys. Rev. Lett.\/} {\bf 99}, 236806
  (2007).

\end{thebibliography}

\textbf{Acknowledgments: }We acknowledge financial support from Microsoft
Corporation Project Q, the NSF GRFP, and the DOE SCGF Program.

\textbf{Author Contributions: }V.V. and S.H. conceived and designed
the experiments, prepared samples, carried out the experiments and
data analysis and wrote the paper. A.Y. conceived and designed the
experiments, carried out data analysis and wrote the paper. L.N.P.
and K.W.W. carried out the molecular beam epitaxy growth.

\textbf{Author Information: }The authors declare no competing financial
interests. Correspondence and requests for materials should be addressed
to yacoby@physics.harvard.edu.

\newpage{}

\begin{figure}[H]
\centering{}\includegraphics[scale=0.4]{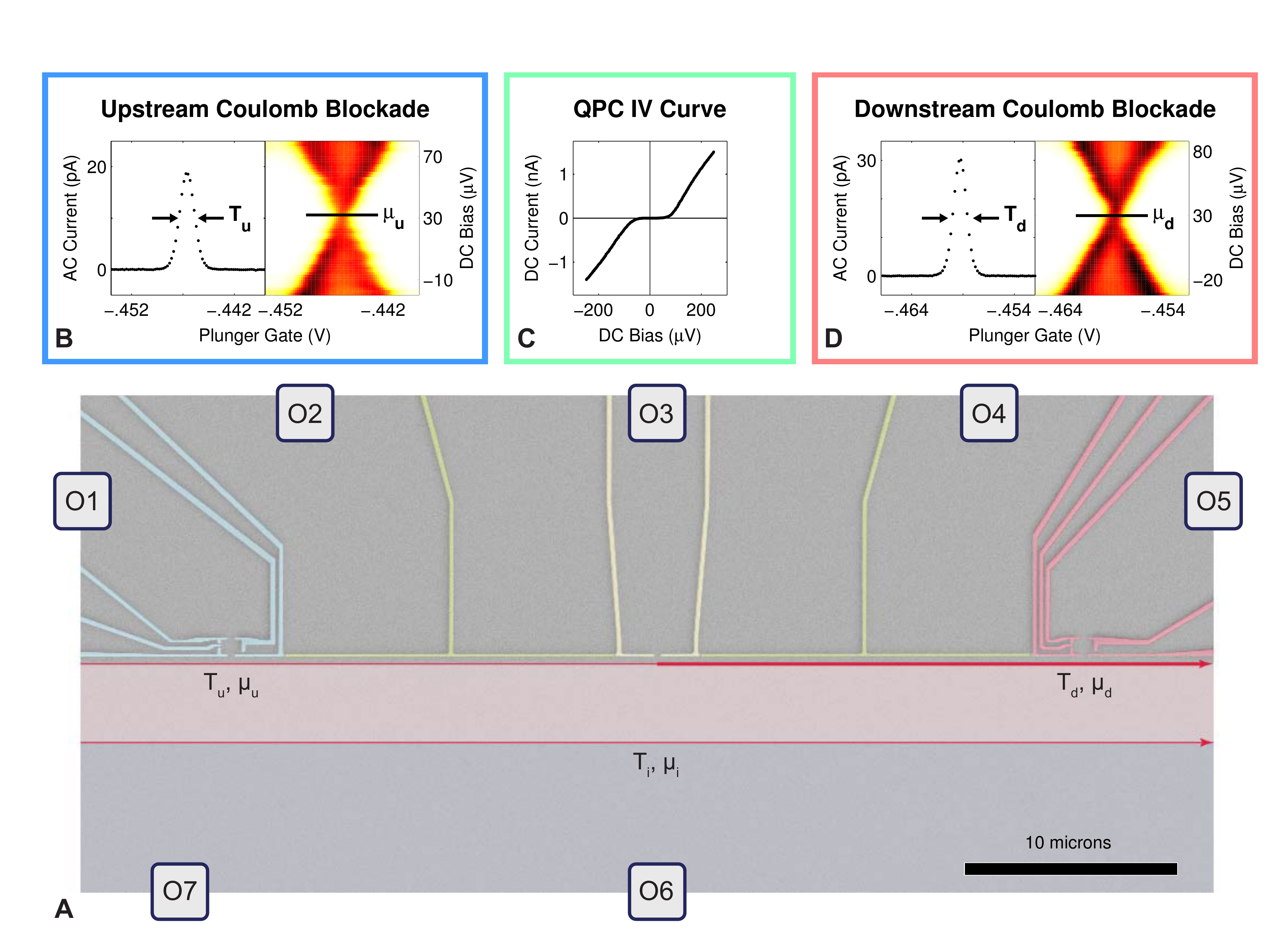}\caption{A) Scanning electron micrograph showing the gate geometry of the device.
O1-7 denote ohmic contacts. Injection of current through the central
quantum point contact (QPC) populates the outermost quantum Hall edge
channel, creating a non-equilibrium distribution. Deflector gates
adjacent to the injection site define the edge or can be de-energized
to deflect edge channels to floating ohmic contacts (O2 and O4). A
quantum dot located 20 microns downstream of the injection site is
used to measure the temperature $T_{d}$ and chemical potential $\mu_{d}$
of the outer edge channel. Similarly, an upstream dot measures $T_{u}$
and $\mu_{u}$. B, D) Coulomb blockade (CB) peaks and diamonds for
the quantum dots. The temperature is determined from the CB peak width.
The chemical potential is determined by zeroing the voltage bias across
the quantum dot. C) The IV characteristic of the QPC.}
\end{figure}

\begin{figure}[H]
\begin{centering}
\includegraphics[scale=0.4]{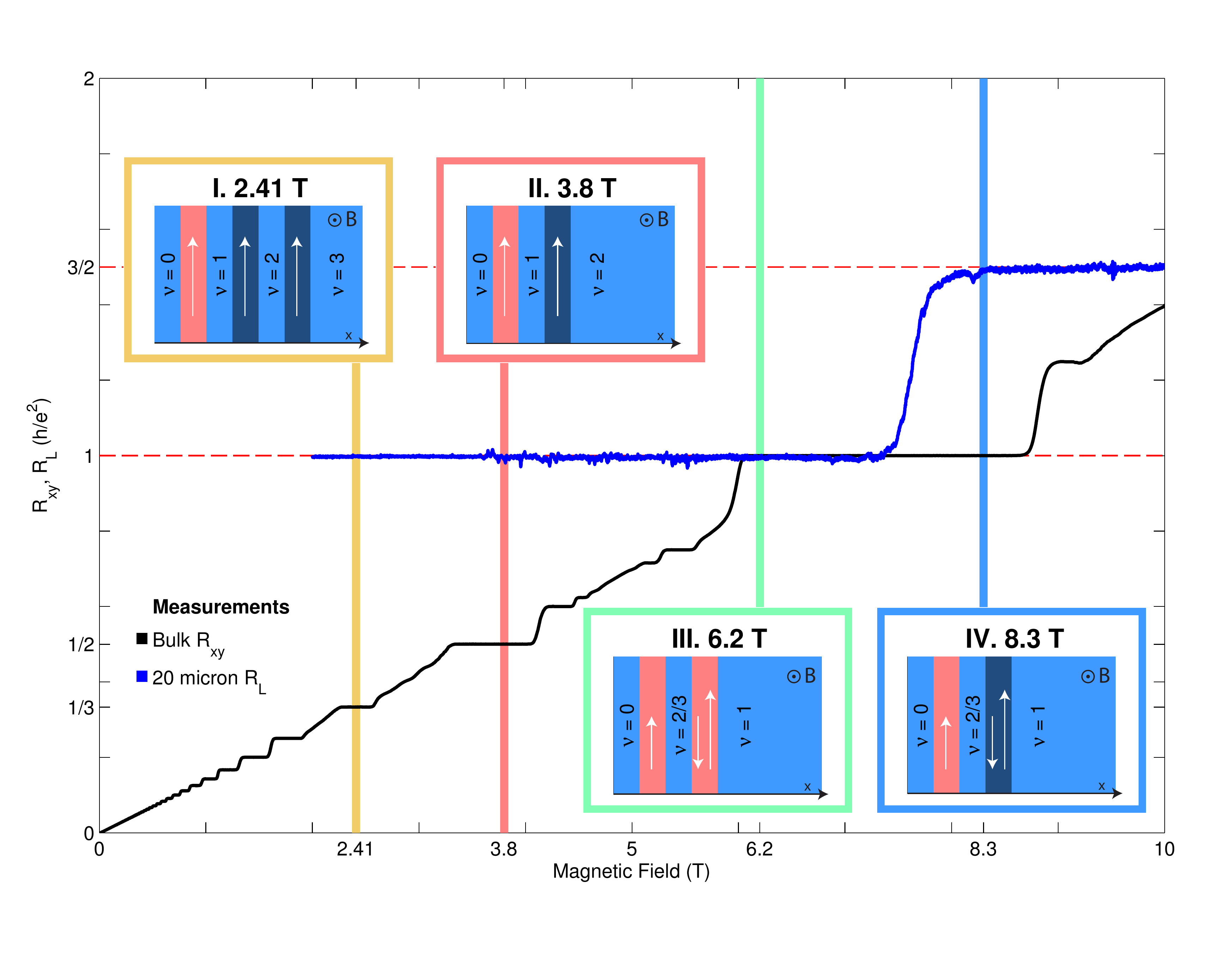}
\end{centering}

\caption{Magnetic field dependence of the Hall resistance $R_{xy}$ (black),
and the local Hall resistance $R_{L}$ (blue). The local Hall resistance
is measured using the central QPC as a current source ($\sim10$ pA)
and a downstream QPC as a voltage probe. Plateaus in $R_{L}$ reveal
the structure of the edge, and also indicate which edge channels participate
in charge transport. The insets depict the qualitative structure of
the sample edge at various magnetic fields, with incompressible regions
shown in light blue and labeled by filling factor. In the intervening
compressible channels, arrows point in the direction of charge flow,
while the arrow length specifies a charge conductance of $G=1$ or
$G=2/3$ in units of $e^{2}/h$. The channels highlighted in red contribute
to charge transport at the voltage probe. I,II) When the bulk filling
factor is $\nu=2$ or $\nu=3$, the edge is composed of integer channels
with the outermost channel having conductance $G=1$. At the voltage
probe, the excess current is carried solely by the outermost channel.
III,IV) Outside the bulk $\nu=1$ state the edge is reconstructed,
resulting in an outermost $G=2/3$ charge channel. The remaining $1/3$
conductance can be found on a spatially separated inner edge located
in the compressible region between the $\nu=2/3$ and $\nu=1$ incompressible
regions. At 8.3 T, the excess current is carried to the voltage probe
only by the outermost channel. At 6.2 T, the edge channels come to
the same potential before reaching the voltage probe, resulting in
$R_{L}=1$.}
\end{figure}

\begin{figure}[H]
\begin{centering}
\includegraphics[scale=0.5]{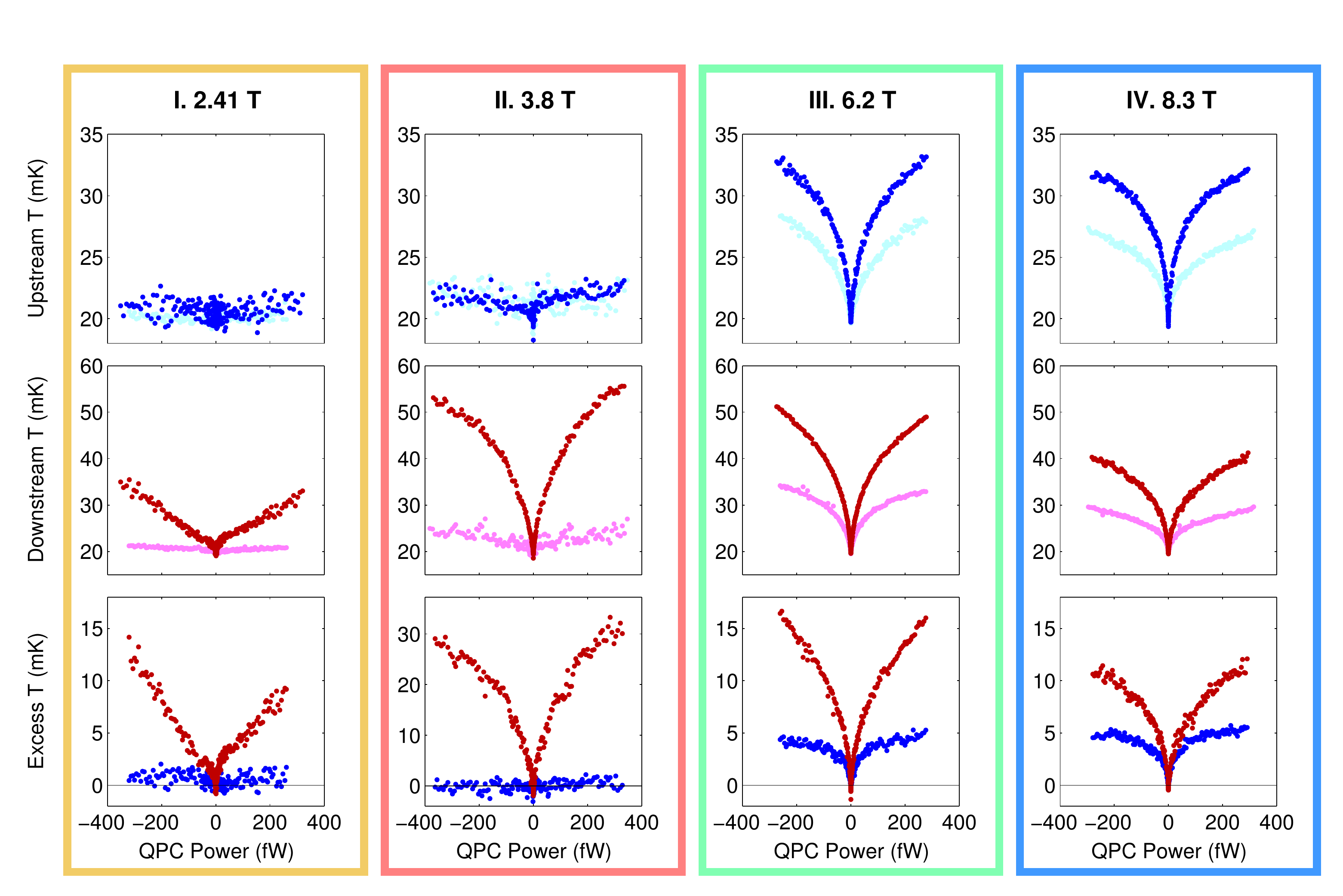}
\end{centering}

\caption{Dependence of the local edge temperature on the power dissipated in
the QPC ($|I|\cdot\Delta V_{QPC})$, at different magnetic fields.
Negative (positive) QPC powers correspond to the injection of holes
(electrons). For each magnetic field, the upstream and downstream
temperatures were measured with (blue, red) and without (cyan, magenta)
the deflector gates energized. With the deflectors at zero voltage
the edge channels are directed to floating ohmic contacts where equilibration
occurs. The difference between temperatures with and without the deflectors
energized, for the same dot, yields the local change in temperature
due to the heat carried by the edge. This excess temperature for the
downstream (red) and upstream (blue) dots is plotted across the bottom
row, for each magnetic field. For I and II, corresponding to an integer
outermost edge, heat is carried chirally downstream with no upstream
heat transport. For IV, where we measure a 2/3 outermost edge, the
heat is carried downstream and upstream. For III, heat is also carried
in both directions, while $R_{L}=1$. We attribute this behavior to
reconstruction outside the bulk $\nu=1$ edge, which allows upstream
heat transport without 2/3 charge transport.}
\end{figure}
\begin{figure}[H]

\centering{}\includegraphics[scale=0.8]{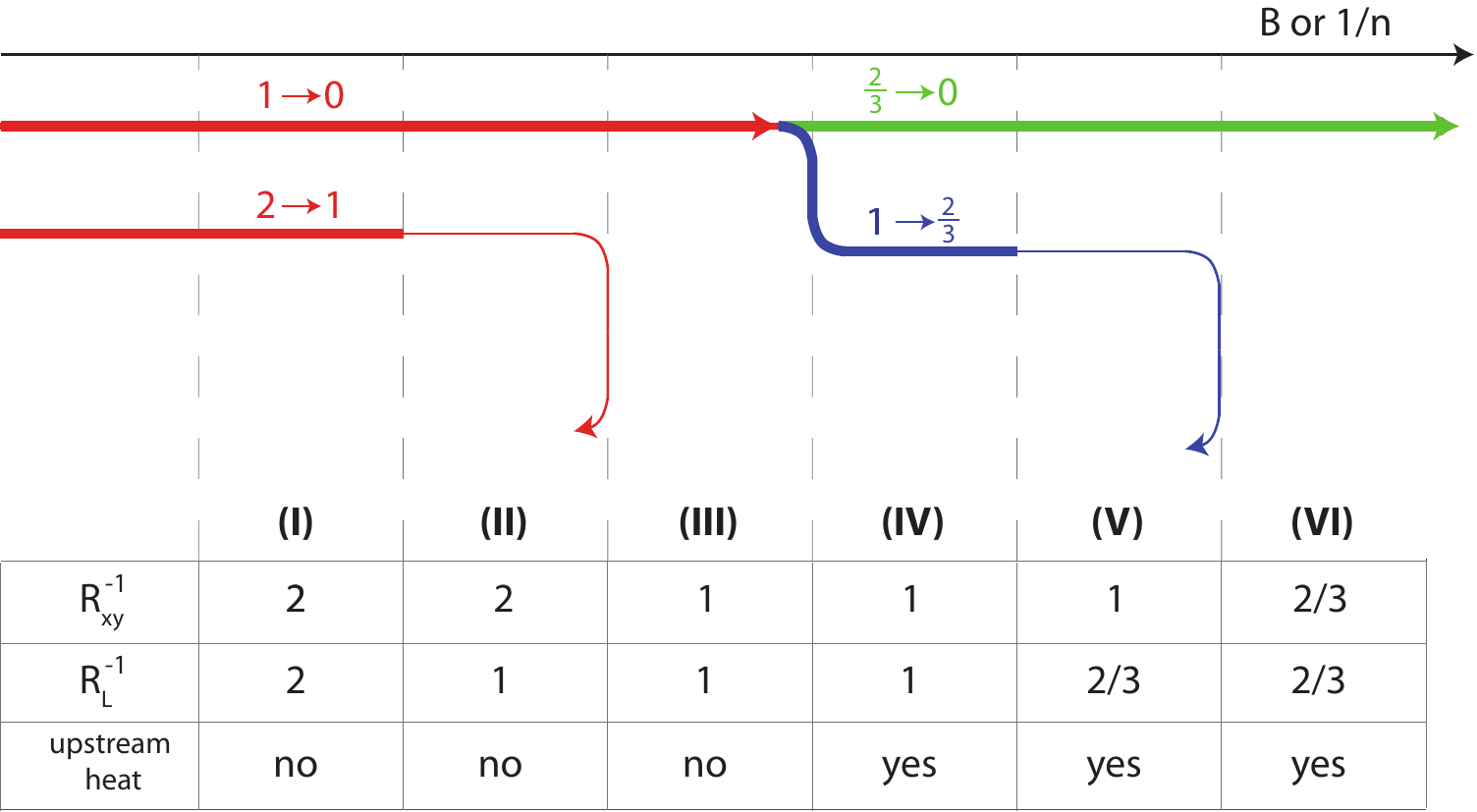}
\caption{One possible evolution of edge structure as a function of magnetic
field. Three types of edges are present in this experiment, denoted
by different colors, with the topmost edge being the outer edge. Only
edge structures II, IV, and V are present in the device in Fig. 1.
Edge III is presented in an online supplement. All resistances are
given in units of $\frac{h}{e^{2}}$. A measurement of $R_{xy}$ allows
one to determine the total conductance of all edges. A measurement
of $R_{L}$ allows one to determine which edges chemically equilibrate
when charge is injected into the outer edge (denoted by bold lines
in the figure). Only the $\frac{2}{3}\rightarrow0$ edge transports
heat upstream, and can be identified by our thermometry measurements.
Detecting upstream heat allows us to discriminate between edges III
and IV (see SOM D).}
\end{figure}

\renewcommand{\thefigure}{\string S\arabic{figure}}
\setcounter{figure}{0}

\newpage{}
\section*{Supplement A: Chemical Potential Measurements and Thermometry}

\subsection*{I. Chemical Potential Measurement }

As the DC current $I$ injected through the QPC increases, the downstream
chemical potential of the outermost edge component must correspondingly
rise. Unless a DC voltage bias $V_{ZB}$ is applied to O5 to exactly
compensate this altered chemical potential, a DC current will flow
through the dot whenever the Coulomb blockade is lifted. Tuning the
dot to this zero bias condition allows us to measure the chemical
potential of the outermost edge component. In principle an upstream
charge current may cause a similar rise in chemical potential at the
upstream dot. For all measurements, the upstream chemical potential
was indistinguishable from that of the ground contact (O7), suggesting
that upstream charge transport does not occur on a 20 $\mu$m scale.

The dependence of $V_{ZB}$ on the current $I$, at a particular value
of magnetic field, measures the total conductance of the edge channels
participating in charge transport at the quantum dot. For the deflector
gates energized, this conductance matches $1/R_{L}$. When the deflector
gates are at zero voltage, however, all edges carry charge and the
total conductance matches the Hall conductance. These observations
corroborate the assertion that the deflector gates are able to direct
the flow of edge channels. When the deflector gates are energized,
the data also show that charge remains in the outermost edge on a
20 $\mu$m scale even during thermometry measurements. An example
of edge resistances determined using the quantum dot zero bias condition
is presented in Figure S1. \medskip{}

\subsection*{II. Coulomb Blockade Thermometry }

At each value of the magnetic field, quantum dots were tuned to the
Coulomb blockade (CB) regime. The typical charging energy was 50 $\mu$eV,
while the typical spacing between CB peaks corresponded to 20 mV on
the plunger gate. We calibrated each dot individually for thermometry
measurements by extracting the slopes $m_{1}$ and $m_{2}$ of CB
diamonds adjacent to the conductance peak of interest, as shown in
Figure S2. The lever arm $\alpha=C_{G}/C$ was then determined by
\begin{equation}
\alpha=\frac{|m_{1}m_{2}|}{|m_{1}|+|m_{2}|},
\end{equation}
where $C_{G}$ is the capacitance between the dot and the plunger
gate, and $C$ is the total capacitance. Knowing $\alpha$ allows
the use of the conductance peak width as a sensitive thermometer.
Our dots are in the metallic regime $\triangle E\ll k_{B}T\ll e^{2}/C$,
where the temperature far exceeds the dot level spacing $\triangle E$.
The temperature of the leads is then found through the formula for
the lineshape of a conductance peak centered at gate voltage $V_{R}$
:
\begin{equation}
G \propto \cosh^{-2}\left(\frac{e\cdot\alpha\cdot|V_{R}-V_{G}|}{2.5k_{B}T}\right).
\end{equation}

During the experiment, we applied a fixed 4 $\mu$V AC voltage bias
and a variable DC voltage bias to each dot (contacts O1 and O5 in
Figure 1). The different AC frequencies used for each dot were typically
215 and 315 Hz. To determine the temperature $T$ of the leads coupled
to a single dot, we first tuned the DC voltage bias $V_{DC}$ applied
to the dot so that the chemical potentials of the two leads were equal,
as described above. Then the plunger gate voltage $V_{G}$ was swept
through a conductance peak while the resulting AC current was monitored
using lockin techniques. The typical AC dot resistance was $>100$
k$\Omega$, resulting in AC currents of $\sim10$ pA. For each DC
current $I$ injected through the QPC, the temperature of the leads
was extracted using equation (2). Representative scans over conductance
peaks in the downstream dot, for two different injected currents,
are plotted in Figure S3.

A data set consisted of one sweep of the DC voltage bias $V_{bias}$
applied to the QPC (contact O3), between -250 $\mu$V and 250 $\mu$V.
At each value of $V_{bias}$ we recorded the injected current $I$,
as well as the temperature $T$ and chemical potential $\mu$ for
both dots. The QPC power was defined as the vector $P_{QPC}=I\cdot V_{bias}-I^{2}(h/\nu e^{2})$,
where $\nu$ was the bulk filling factor. For each sweep, the electron
temperatures found using equation (2) were normalized such that the
minimum electron temperature was always 20 mK, equivalent to an effective
rescaling of $\alpha$. This minimum electron temperature of 20 mK
was measured at the base temperature of our dilution refrigerator
via Coulomb blockade thermometry, for quantum dots with cold leads
sourced directly from ohmic contacts. We assume in our experiment
that all edges are at this minimum temperature when $V_{bias}=0$.
For a dot coupled to a fractional edge, electronic correlations may
alter the temperature extracted using equation (2). As long as the
peak width remains linear in temperature as a result of such behavior,
our procedure accurately reports relative edge temperatures. The absolute
fractional edge temperatures may then differ from our reported data
by an overall normalization.

While all of our reported Coulomb blockade temperatures use the above
rescaling to normalize the base temperature to 20 mK, it is also possible
to calibrate temperatures using the resistive RuO thermometer on the
mixing chamber. In Figure S4 such a calibration is plotted, showing
how the temperature deduced from Coulomb blockade peaks corresponds
to the mixing chamber temperature. The behavior is linear at high
temperatures and saturates to the minimum dot temperature of 20 mK
at low temperatures due to the decoupling of the electronic system
from the lattice. Because the mixing chamber thermometer is not directly
coupled to the two-dimensional electronic system, we have chosen to
normalize minimum temperatures to 20 mK rather than calibrate using
the mixing chamber. From this data we see that a calibration using
the mixing chamber thermometer does not significantly alter our results
beyond a \textasciitilde{}140\% temperature rescaling at the highest
reported temperatures. None of our qualitative claims are changed
by such a rescaling, and this temperature increase cannot explain
the temperature deficiency discussed in the main text and Supplement
B. 

We accumulated several normalized data sets at each value of magnetic
field, both with the deflector gates energized and at zero voltage.
To determine the increase in temperature at the downstream dot due
to heat carried by the outermost edge component, we first separated
the data sets into two groups, depending on whether the deflectors
were energized or at zero voltage while the data was taken. For each
group, the normalized downstream temperatures were then averaged to
obtain two vectors containing the mean downstream temperatures for
both deflector settings. The QPC powers were similarly averaged, resulting
in the power-dependent mean temperatures plotted in Figure 3. The
difference between the mean downstream temperatures, for equal QPC
power, was reported as the excess downstream dot temperature. This
procedure was also used for the upstream dot, and for all reported
values of the magnetic field. The excess temperatures determined in
this way are plotted in the third row of Figure 3 of the main paper.

\begin{figure}
\begin{centering}
\includegraphics[scale=0.5]{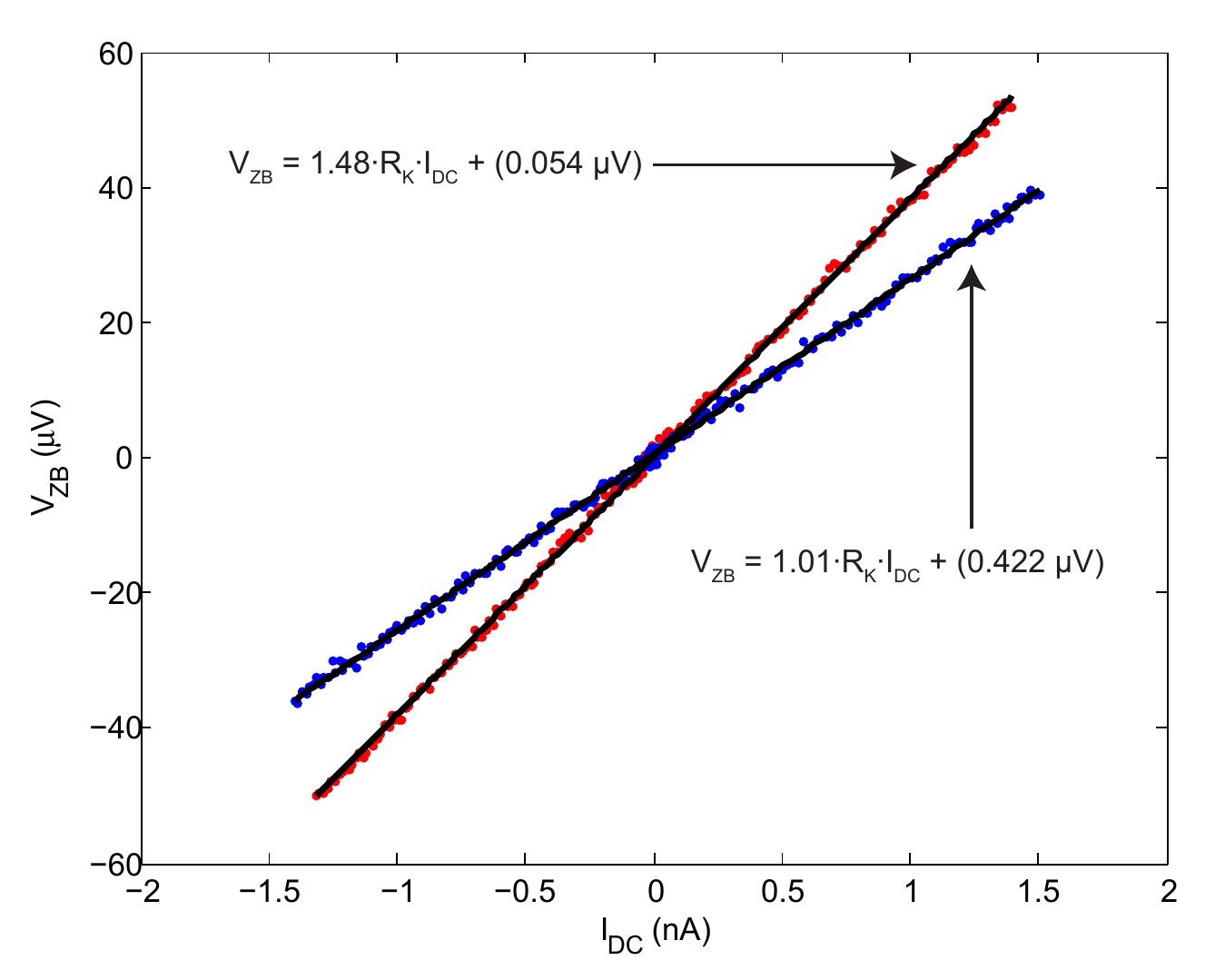}
\par\end{centering}

\caption{The voltage $V_{ZB}$ applied to O5 in order to zero the bias across
the downstream quantum dot, as a function of the current $I_{DC}$
injected through the QPC. Data were acquired at a magnetic field of
8.3 T. For the deflector gates energized, $V_{ZB}$ is shown in red,
with a slope corresponding to current carried by a $\nu=2/3$ outermost
edge. When the deflector gates are set to zero, $V_{ZB}$ is shown
in blue, indicating conduction of current by edges with total conductance
$G=1$.}

\end{figure}

\begin{figure}
\begin{centering}
\includegraphics[scale=0.7]{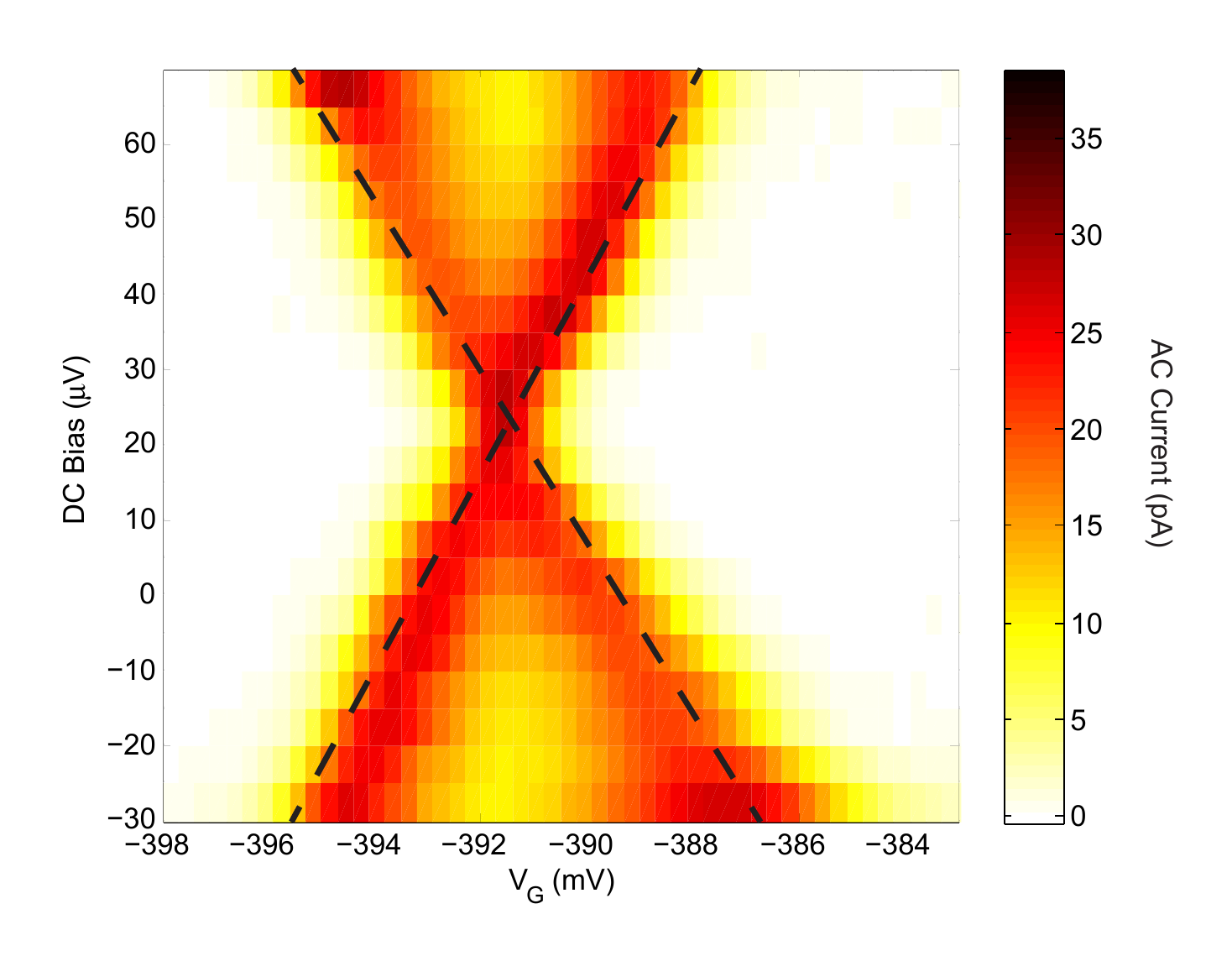}
\par\end{centering}

\caption{Coulomb blockade data used to calibrate the downstream quantum dot
at a magnetic field of 8.3 T. The lever arm $\alpha$ is calculated
from the slopes of the zero-conductance regions, marked by dashed
lines.}
\end{figure}

\begin{figure}
\begin{centering}
\includegraphics[scale=0.7]{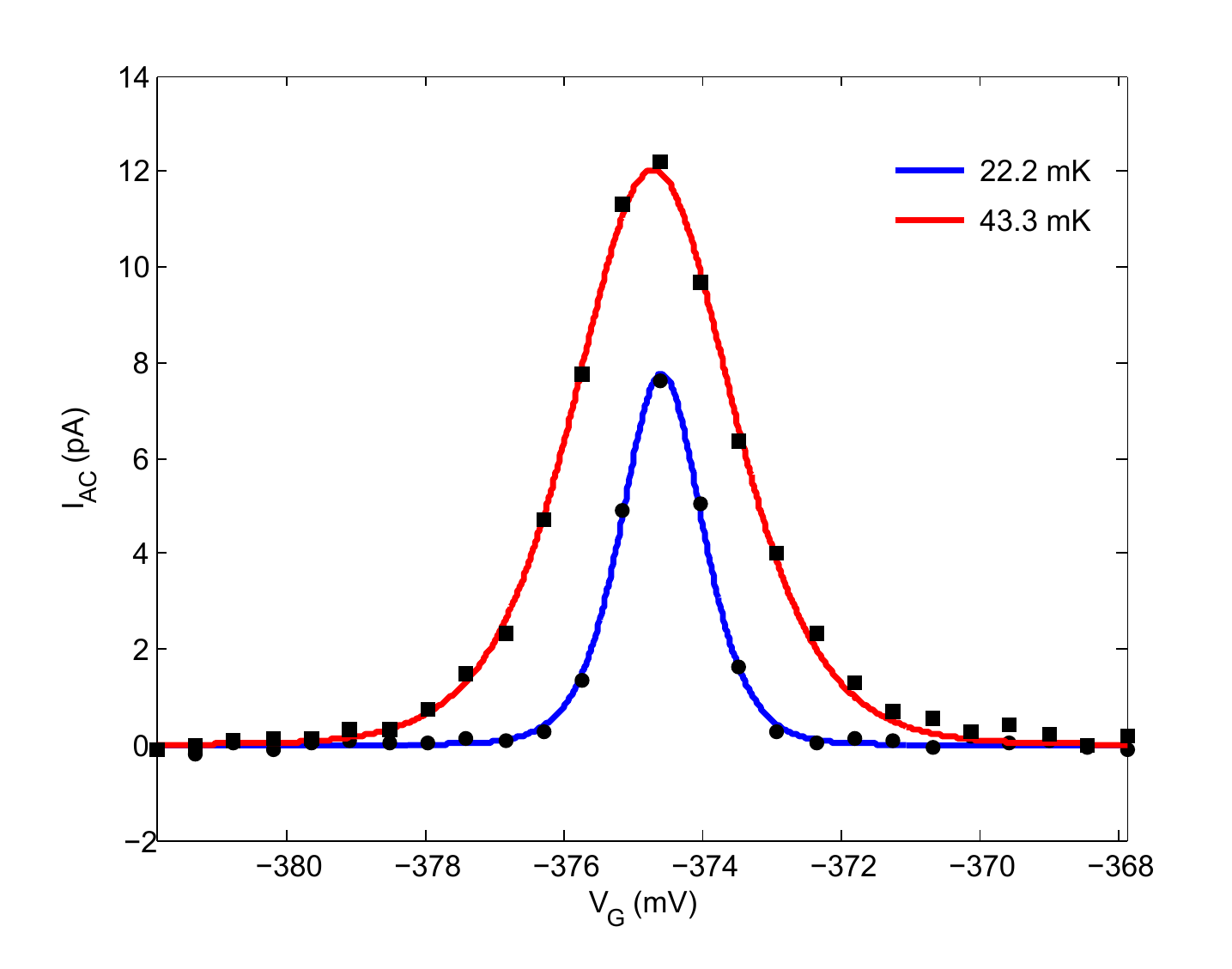}

\par\end{centering}

\caption{AC current $I_{AC}$ through the downstream quantum dot as a function
of the plunger gate voltage $V_{G}$, measured at a magnetic field
of 8.3 T. Black circles (squares) correspond to data taken for an
injected QPC current $I=0$ nA ($I=1.5$ nA). The best fits of the
data to equation (2) are shown in blue and red, and give temperatures
of $T=22.2$ mK for $I=0$ nA and 43.3 mK for $I=1.5$ nA.}

\end{figure}

\begin{figure}[H]
\begin{centering}
\includegraphics[scale=0.4]{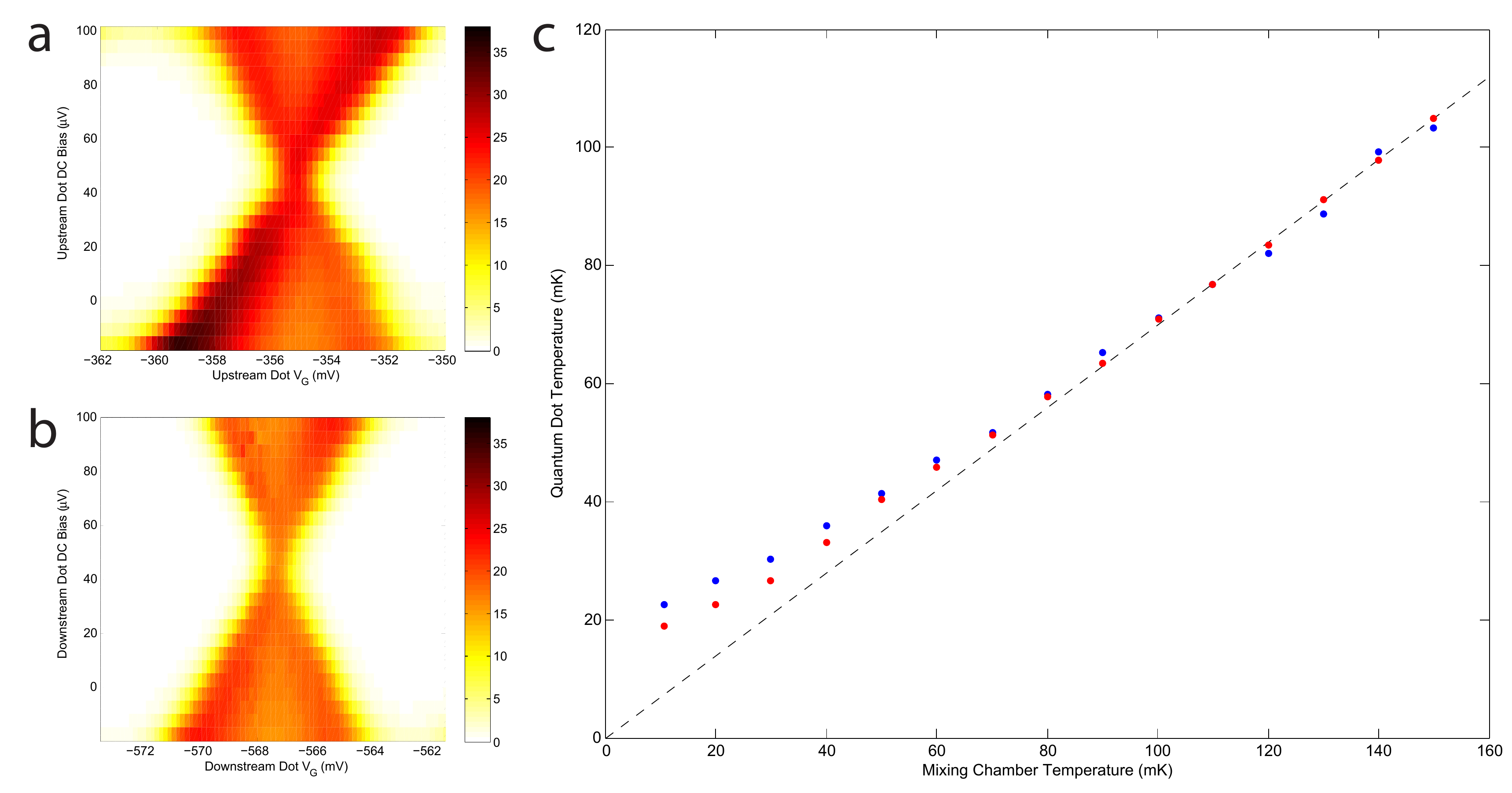}
\end{centering}

\caption{Comparison of Mixing Chamber and Coulomb Blockade Thermometry, at
6.15 T. Heat is applied to the mixing chamber and temperatures are
measured using a resistive RuO thermometer attached to the mixing
chamber along with the two patterned quantum dots. a,b) Widths for
the CB thermometers are calibrated using the diamonds. c) The widths
of our CB peaks are linear in temperature, except for a saturation
at 20 mK as the mixing chamber is cooled to 10 mK.}

\end{figure}

\section*{\pagebreak{}}

\section*{Supplement B: Power carried by Non-Equilibrium Edges}

In our experiment, we tune the bulk quantum Hall state to filling
factor $\nu$, and apply a voltage $V$ between ohmic contacts O3
and O6. These two contacts are separated by a QPC tuned to have resistance
$R\sim100$ k$\Omega$. When a net current $I$ is injected through
the QPC, the electronic occupation of the outermost compressible edge
channel deviates locally from its equilibrium distribution. Quantum
dots placed 20 $\mu$m upstream and downstream of the QPC probe the
chemical potential and temperature of this outermost edge. The form
of Coulomb blockade peaks monitored during our heat transport measurements
suggests that the outer edge internally reaches thermal equilibrium
over a distance smaller than 20 $\mu$m. However, our charge measurements
indicate that chemical equilibration of the outer edge with inner
edge channels starts to occur at a distance greater than 20 $\mu$m.
Thus, at the downstream measurement point the outermost edge has a
Fermi occupation function and carries all of the injected current
$I$. For the measurements at magnetic fields of 2.41 T and 3.8 T
(bulk $\nu$ = 3 and $\nu$ = 2), the electrical conductance of this
edge is consistent with downstream charge transport by a single integer
quantum Hall (IQH) edge. Furthermore, our thermometry measurements
show strict downstream heat transport, also consistent with the IQH
regime. To determine the expected quantitative outcome of our measurements
in the IQH regime, we analyze charge and heat transport by IQH edges
in the experimental system described above.

The chemical potential $\mu$ of an IQH edge is related to the current
$I_{E}$ that it carries:
\begin{equation}
I_{E}=\frac{e}{h}\mu.
\end{equation}
In our model, the total number of edge channels on each side of the
QPC is equal to the bulk filling $\nu$. However, since only the outermost
channel contributes to charge transport on a 20 $\mu$m scale, we
treat inner channels as inert and consider only the behavior of the
outer channel. The two outer edges that carry charge toward the QPC
originate in ohmic contacts O3 and O6. The occupations of these incoming
edges are therefore Fermi functions,
\begin{equation}
\begin{aligned}f_{in}^{O3}(E) & =f(E-\mu_{in}^{O3},T_{base})\\
f_{in}^{O6}(E) & =f(E-\mu_{in}^{O6},T_{base}),
\end{aligned}
\end{equation}
where $\mu_{in}^{O3}=\mu+eV$ and $\mu_{in}^{O6}=\mu$ are the chemical
potentials of O3 and O6 and $T_{base}$ = 20 mK is the electron base
temperature. At the QPC, the electronic occupations of the outgoing
edge modes are forced out of equilibrium. At a distance 20 $\mu$m
from the QPC these outgoing edges reach thermal equilibrium, with
chemical potentials $\mu_{out}^{O3}=\mu+eV-(h/e)I$ and $\mu_{out}^{O6}=\mu+(h/e)I$
determined using equation (3). While these chemical potentials can
be found simply by considering charge transport, a more detailed analysis
of scattering at the QPC is necessary to determine the temperatures
of the outgoing edges.

The equilibrium temperature $T$ of an IQH edge is related to the
power $J_{E}$ carried by its excitations according to
\begin{equation}
J_{E}=\frac{(\pi k_{B})^{2}}{6h}T^{2}.
\end{equation}
In general $J_{E}$ can also be calculated from the occupation $n(E)$
and chemical potential $\mu$ of an edge, by integrating the power:
\begin{equation}
J_{E}=\frac{1}{h}\int_{0}^{\mu}dE\cdot(\mu-E)\cdot(1-n(E))+\frac{1}{h}\int_{\mu}^{\infty}dE\cdot(E-\mu)\cdot n(E).
\end{equation}
Here the first integral corresponds to the contribution of hole-like
excitations, while the second integral corresponds to particle-like
excitations. The 1D relation $g(E)\cdot v(E)=1/h$ between the velocity
$v(E)$ and density of states $g(E)$ was used to simplify the integrals.

Since the outgoing edges in our model have non-equilibrium distributions
$n_{out}^{O3}(E)$ and $n_{out}^{O6}(E)$ immediately after the injection
of current $I$, their respective energy currents are determined using
equation (6). At a distance 20 $\mu$m from the QPC, the outgoing
edges are in equilibrium. If no energy escapes from the edge as it
equilibrates, equation (5) then provides a calculation of the expected
edge temperatures. With the goal of ultimately finding these temperatures,
we therefore consider the forms of the non-equilibrium edge distributions,
which depend on the energy-dependent QPC transmission probability
$\tau(E)$. This transmission is determined by the differential conductance
$dI/dV$ of the QPC, as follows:
\begin{equation}
I=\int_{0}^{\infty}dE\cdot\tau(E)\cdot(f_{in}^{O3}(E)-f_{in}^{O6}(E)).
\end{equation}

Using $\tau$ and the distributions of the incoming edges (4), we
find expressions for the non-equilibrium distributions: 
\begin{equation}
\begin{aligned}n_{out}^{O3} & =(1-\tau)\cdot f_{in}^{O3}+\tau\cdot f_{in}^{O6}\\
n_{out}^{O6} & =(1-\tau)\cdot f_{in}^{O6}+\tau\cdot f_{in}^{O3}.
\end{aligned}
\end{equation}
From these distributions we can then deduce the partitioning of power
among the outgoing edges, as well as the outgoing equilibrium temperatures
$T_{out}^{O3}$ and $T_{out}^{O6}$. We find that each outgoing edge
carries an equal energy current. Conservation of energy provides a
constraint on the total outgoing power: 
\begin{equation}
I\cdot V-I^{2}(h/e^{2})=\frac{(\pi k_{B}T_{out}^{O3})^{2}}{6h}+\frac{(\pi k_{B}T_{out}^{O6})^{2}}{6h}-\frac{(\pi k_{B}T_{base})^{2}}{3h}.
\end{equation}
This relationship holds as long as the inner edges remain decoupled
from the outermost edge modes. Here the left-hand side specifies the
power dissipated by the QPC, while the right-hand side represents
the net power carried away by edge excitations. The term $I^{2}(h/e^{2})$
refers to energy dissipated at ohmic contacts, and does not contribute
to heating the edge. For completeness, the distributions of the outgoing
edges, 20 $\mu$m from the QPC, are given below:
\begin{equation}
\begin{aligned}f_{out}^{O3}(E) & =f(E-\mu_{out}^{O3},T_{out}^{O3})\\
f_{out}^{O6}(E) & =f(E-\mu_{out}^{O6},T_{out}^{O6}),
\end{aligned}
\end{equation}

In Figure S5, numerical calculations of the outermost edge occupation
functions are plotted during each stage of scattering at the QPC,
for an applied voltage $V=175$ $\mu$V and at bulk filling $\nu=2$.
In panel B, the incoming distributions are shown with the QPC transmission
$\tau$ extracted from IV data. In panels C and D, the non-equilibrium
and equilibrium distributions are plotted for outgoing edges on each
side of the QPC. For the equilibrium outgoing distributions, we extract
the temperatures $T_{out}^{O3}$ and $T_{out}^{O6}$ over a range
of $V$ to determine the dependence of edge temperatures on the QPC
power $P_{QPC}=I\cdot V-I^{2}(h/e^{2})$. As shown in Figure S6, our
model qualitatively explains the cusp in temperature that is observed
at $P_{QPC}=0$.

Using this model, we expect the downstream quantum dot to measure
a maximum temperature of $560$ mK for $\nu$ = 2 and $545$ mK for
$\nu$ = 3. The actual observed maximum temperatures were $55$ mK
for $\nu$ = 2 and $35$ mK for $\nu$ = 3. Although we observe no
charge leakage to inner edge channels on a 20 $\mu$m scale, the loss
of heat to inner edges is still possible and would decrease the expected
temperatures. If all edges equilibrate thermally over a distance smaller
than 20 $\mu$m, we expect that the power $J_{E}$ carried by the
outermost edge will be divided by the filling factor $\nu.$ Using
equation (5), it follows that the temperature will be divided by $\nu^{1/2}$.
For this type of thermal equilibration we thus expect to measure $395$
mK for $\nu$ = 2 and $315$ mK for $\nu$ = 3. Whether or not heat
escapes to the inner edges, it is still clear from this analysis that
in our experiment the majority of the power dissipated in the QPC
does not find its way to the outermost edge.
\begin{figure}
\begin{centering}
\includegraphics[scale=0.45]{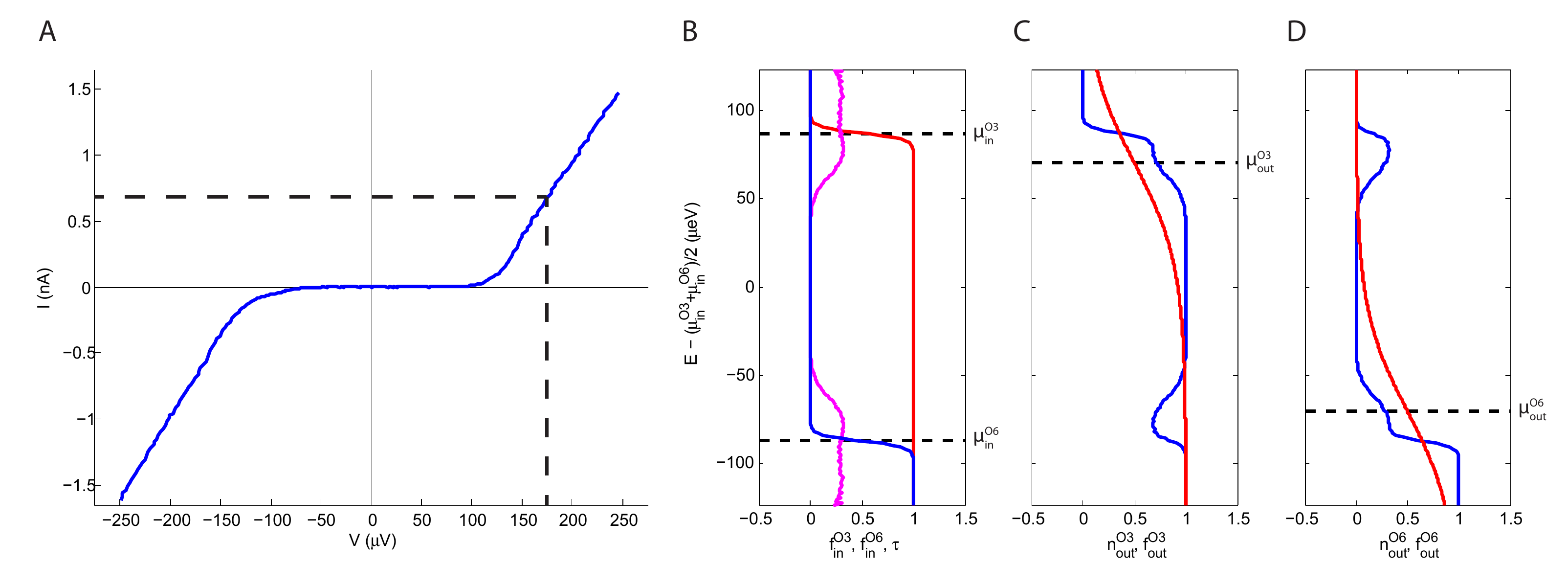}
\par\end{centering}

\caption{A) The IV curve of the QPC at bulk filling $\nu=2$. An applied voltage
$V=175$ $\mu$V was used to calculate the distributions shown in
(B-D). This voltage and the corresponding injected current are marked
with dashed lines. B) The QPC transmission probability $\tau$, calculated
from the QPC IV curve, is shown in magenta. In blue (red), the occupation
$f_{in}^{O6}$ ($f_{in}^{O3}$) of the incoming outer edge mode originating
at ohmic contact O6 (O3). The chemical potentials differ by $eV=175$
$\mu$eV. C) In blue, the non-equilibrium occupation $n_{out}^{O3}$
of the outermost edge immediately after the injection of current through
the QPC. This edge component carries charge toward O3. 20 $\mu$m
from the QPC, the edge is in equilibrium with the distribution $f_{out}^{O3}$,
shown in red. C) The edge component carrying charge toward O6 has
the non-equilibrium occupation $n_{out}^{O6}$, shown in blue, immediately
after current is injected. 20 $\mu$m downstream the edge has equilibrated
to the distribution $f_{out}^{O6}$, shown in red.}
\end{figure}
\begin{figure}[H]
\begin{centering}
\includegraphics[scale=0.5]{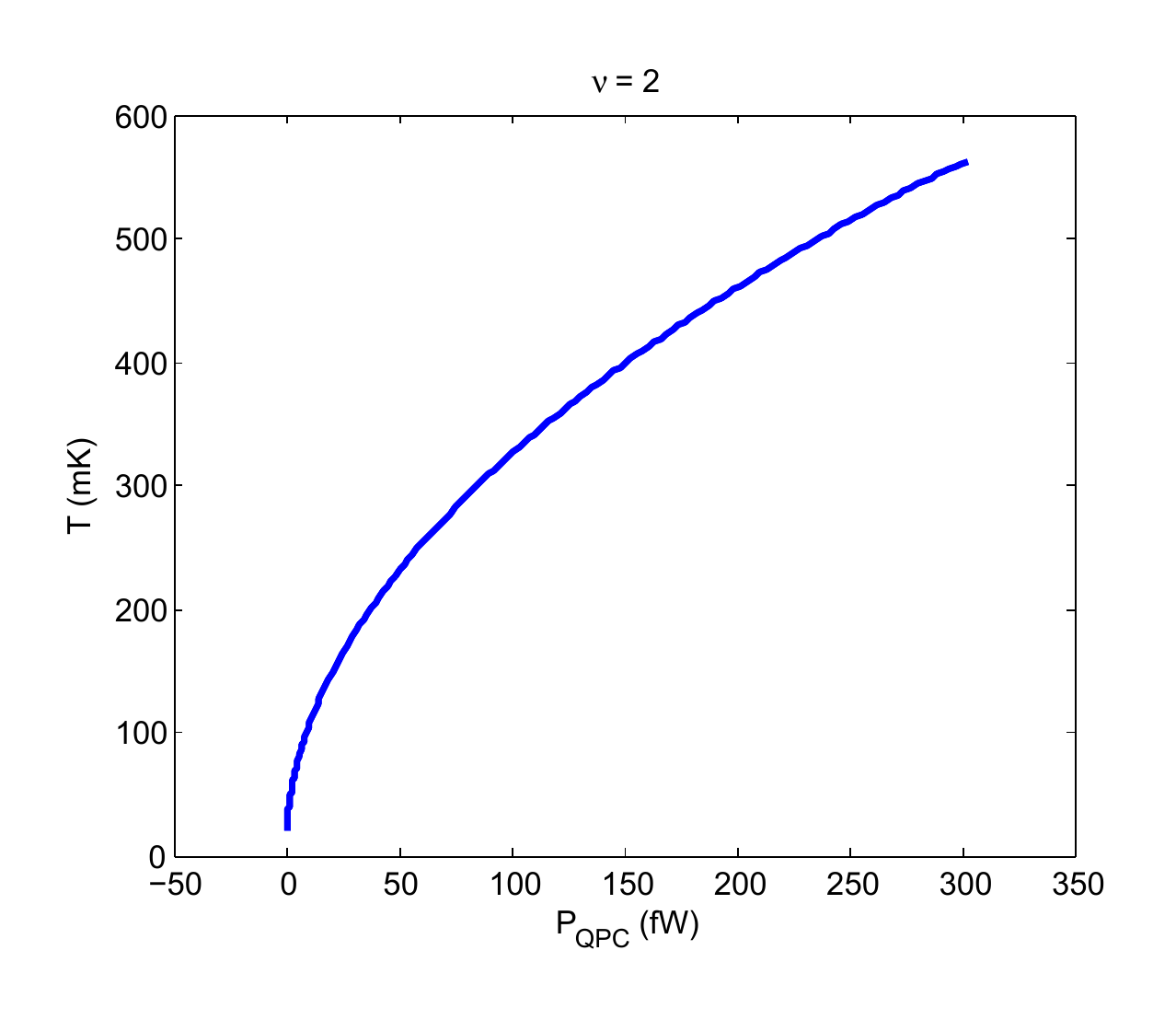}
\par\end{centering}

\caption{Expected equilibrium temperatures of the outgoing outermost edges,
calculated using the measured QPC transmission $\tau$ at $\nu$ =
2. Both outgoing edges are expected to have the same temperature.
The $\nu$ = 3 data give the same temperature as a function of QPC
power.}
\end{figure}

\section*{\pagebreak{}}

\section*{Supplement C: Bulk Heat Transport}

As mentioned in the main text and Supplement B, we observe temperatures
well below what is expected for a system of quantum Hall edges with
no energy dissipation. This necessarily means that heat diffuses out
of the edges into additional modes in either the bulk of the 2D electronic
system or the surrounding crystalline solid. Because we see a bulk
contribution to heating when the bulk is at $\nu=1$ (Columns III
and IV of Figure 3), but not when the bulk is at \textbf{$\nu=2$
}or $\nu=3$ (Columns I and II), and because we don't expect a change
of magnetic field to significantly affect heat conduction through
the solid, we can attribute the heating at high fields to the $\nu=1$
electronic system. While we don't know the mechanism of this bulk
heat transport in such a strongly insulating state, we suspect it
may be associated with low energy spin degrees of freedom that exist
at $\nu=1$.

The presence of this bulk heat transport in the two measurements where
we see upstream heat transport attributed to edges requires some additional
discussion.%
\footnote{A measurement where bulk heat transport is present without edge heating
is presented in Supplement D%
} Specifically, we need to rule out the possibility that turning our
deflectors on and off affects the quantity of heat transported by
the bulk to the thermometers, thereby producing a signal unrelated
to edge heat transport. Below we describe two experiments specifically
designed to address this possibility. Our findings provide two important
observations. Firstly, our gates are not completely effective at preventing
the flow of heat. We inferred this from the shape of our Coulomb blockade
peaks, and checked it explicitly by attempting to block heat flow
with a gate. Secondly, if we reduce the length of the deflector gates
to the point where there is much less bulk $\nu=1$ region for heat
to diffuse upwards into when the deflectors are off, we observe the
same qualitative and quantitative behavior that was presented in the
main body of the paper. Both of these observations are discussed more
carefully below.\medskip{}

\subsection*{I. Diffusion of heat through gated regions}

When our topgates are energized to completely deplete carriers from
the underlying 2D electron system, we would expect that energy can
no longer be transported by that system. However, heat that manages
to diffuse into the lattice can still propagate. Here we will present
data suggesting that some heat does indeed diffuse across the depleted
regions. 

The first indication of such diffusion is taking place can be seen
in the form of our Coulomb blockade peaks. The fits we used in the
experiment assume that the temperatures of the two quantum dot leads
are identical. However, since we are only explicitly heating one side
of the dot, a simple model suggests that we should expect leads with
different temperatures. This temperature asymmetry should show up
as an increased kurtosis in the CB peak shape. In Figure S7, we show
the one-temperature fit that was used in the main body of this paper
along with two alternatives that allow for asymmetric lead temperatures.
This particular peak corresponds to the downstream measurement at
a magnetic field of 6.2 T and an injected power of 274 fW. The deflector
gates are energized, so this peak includes both edge and bulk contributions. 

Figure S7b presents an alternative fit with an additional fit parameter
that allows for different temperatures in the two leads. While the
one-temperature fit suggests lead temperatures of 51 mK, the two temperature
fit suggests that one of the leads is hotter (60 mK) and the other
is colder (39 mK). However, even though the fits are consistently
better with the extra parameter, the residuals are not systematically
cleaner. Figure S8 presents a comparison of the one-temperature and
two-temperature fits for the entire range of injected powers that
we studied. Below 50 fW of injected power, the one-temperature and
two-temperature fits agree exactly, suggesting equal temperature leads.
At higher powers, the two temperature fit does suggest a difference
in the lead temperatures. Even this asymmetry, however, has to be
considered carefully. Because there are nearby peaks (roughly 800
mK away from the peak center, when translated from gate voltage as
in Figure S7), at high temperatures we can expect them to artificially
distort our peak and increase the quality of an asymmetric-temperature
fit. 

Figure S7c presents yet another fit which assumes that the cold lead
has the naively expected temperature of 20 mK, corresponding to the
observed base temperature for electrons with no intentional heating.
The temperature of the hot lead is allowed to vary. With this constraint,
the best fit suggests a hot lead temperature of 59 mK. Here, however,
the residuals have a pronounced trend that persists for all fits with
the 20 mK constraint. 

Without strong evidence that a two-temperature fit better describes
our measurements, we opted to use a single-temperature fit for the
main data presented. None of the qualitative observations of bulk
heat transport or upstream heat transport by a neutral edge mode are
affected by using the hotter temperature from two-temperature fits.
Furthermore, the two-temperature fit doesn't solve the temperature
deficiency alluded to in the main text or Supplement B.

\begin{figure}[H]
\begin{centering}
\includegraphics[scale=0.55]{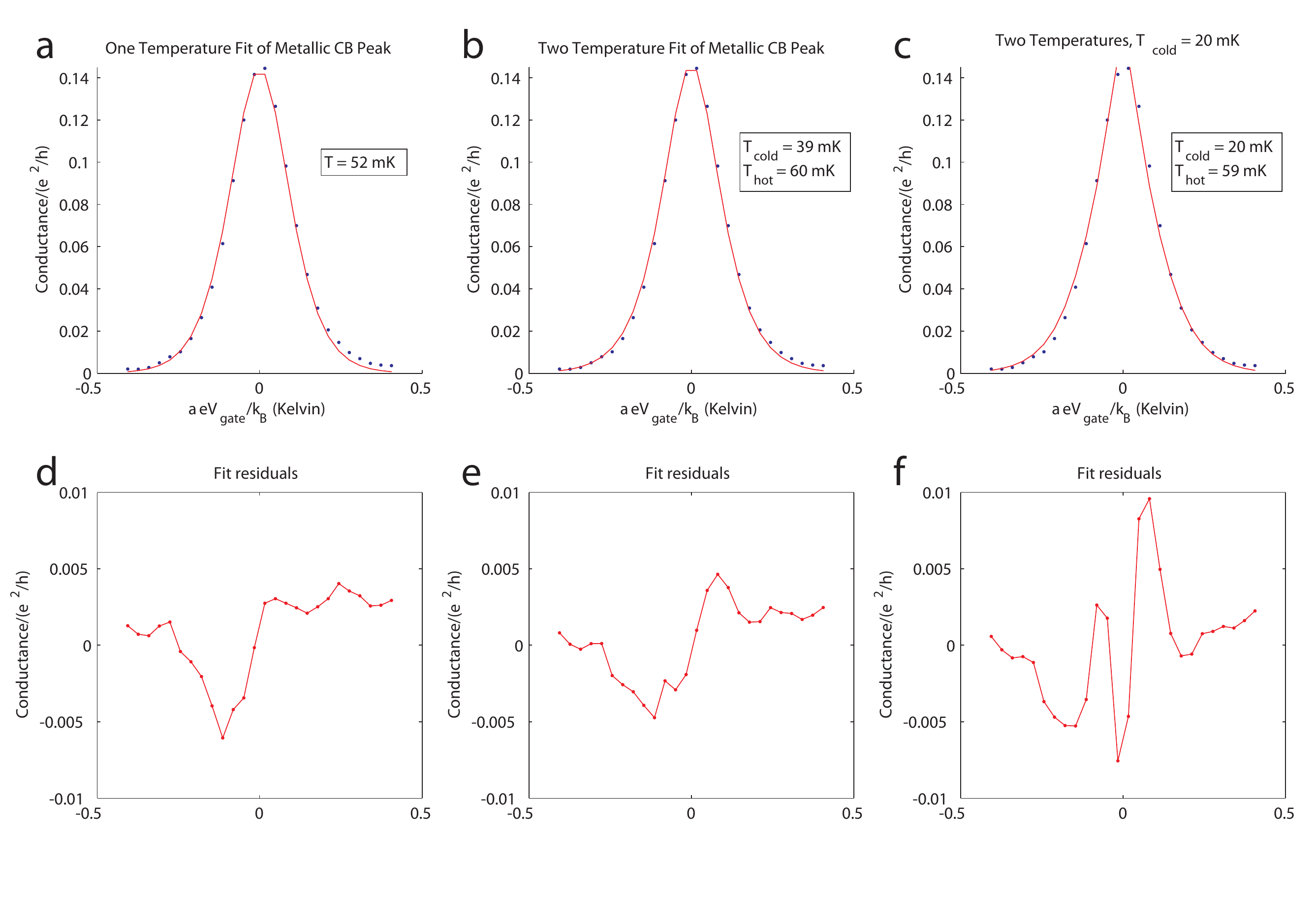}
\par\end{centering}

\caption{\textbf{a)} One temperature fit of the CB peak seen downstream at
6.2T and 274 mW of injected power. The deflector gates are energized,
so this peak includes both edge and bulk contributions. Additional
peaks are centered roughly 800 mK to the left and right of the center
of this peak. \textbf{b) }Fit obtained by adding an additional parameter
allowing for asymmetric lead temperatures. There is no systematic
improvement in the residual trend by using such a fit (though the
quality of fit obviously improves slightly). \textbf{c)} Fit obtained
using the same form as panel b, but fixing the cold lead to a 20 mK
distribution. This produces a low quality of fit and certainly doesn't
describe our data well. \textbf{d,e,f)} Fit residuals plotted below
the associated fit}
\end{figure}
\begin{figure}[H]
\begin{centering}
\includegraphics[scale=0.7]{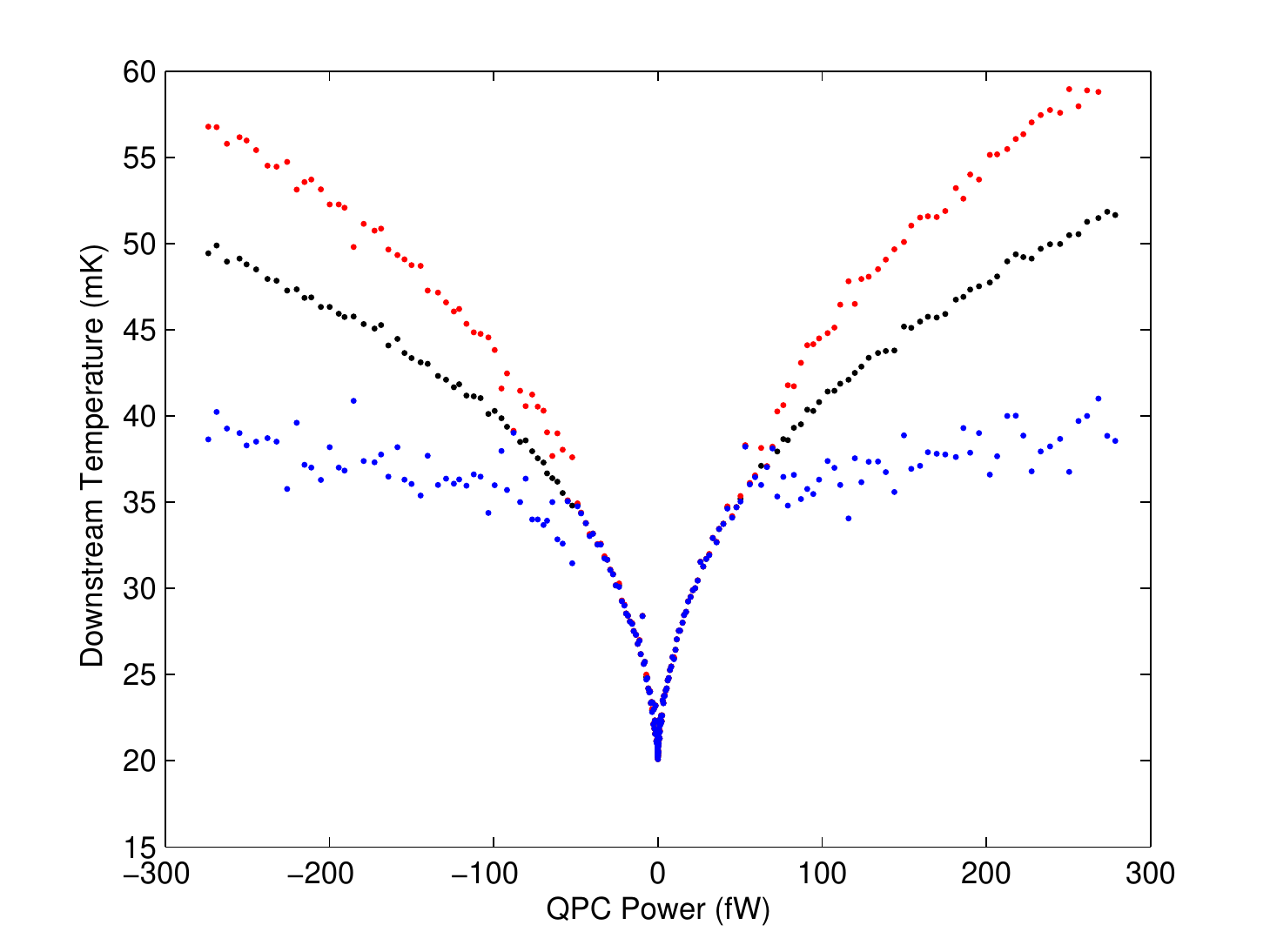}
\par\end{centering}

\caption{Temperature fit of the CB peak seen downstream at 6.2T as a function
of injected power. Black denotes the one-temperature fit (as in Figure
S7a), and red and blue denote the hot and cold temperatures of a two-temperature
fit (as in Figure S7b). They agree perfectly at low injected powers,
but begin to diverge beyond 50 fW. }
\end{figure}

We can go further and explicitly test for heat transport across depleted
regions by placing a strip of such a region between our heater and
our thermometers, as in the device pictured in Figure S9. Any heat
detected at the thermometers would necessarily have to diffuse through
depleted region beneath the vertical gates. Results of this test are
depicted in Figure S10. Here, we can clearly see that some heat flows
through these narrow depleted regions. At the highest injected power,
we see the temperature rise from 20 mK to 28 mK with an uninterrupted
2D and a temperature rise to 22 mK with the 2D depleted beneath the
vertical gates. This small heat diffusion through gated regions is
qualitatively consistent with our observation of heating in the cold
leads of our quantum dots, as mentioned above. The fact that the temperature
is reduced from the ungated value (22 mK versus 28 mK) provides additional
evidence that the 2D electron system is responsible for the observed
bulk heat transport at high fields.

\begin{figure}[H]
\begin{centering}
\includegraphics[scale=0.4]{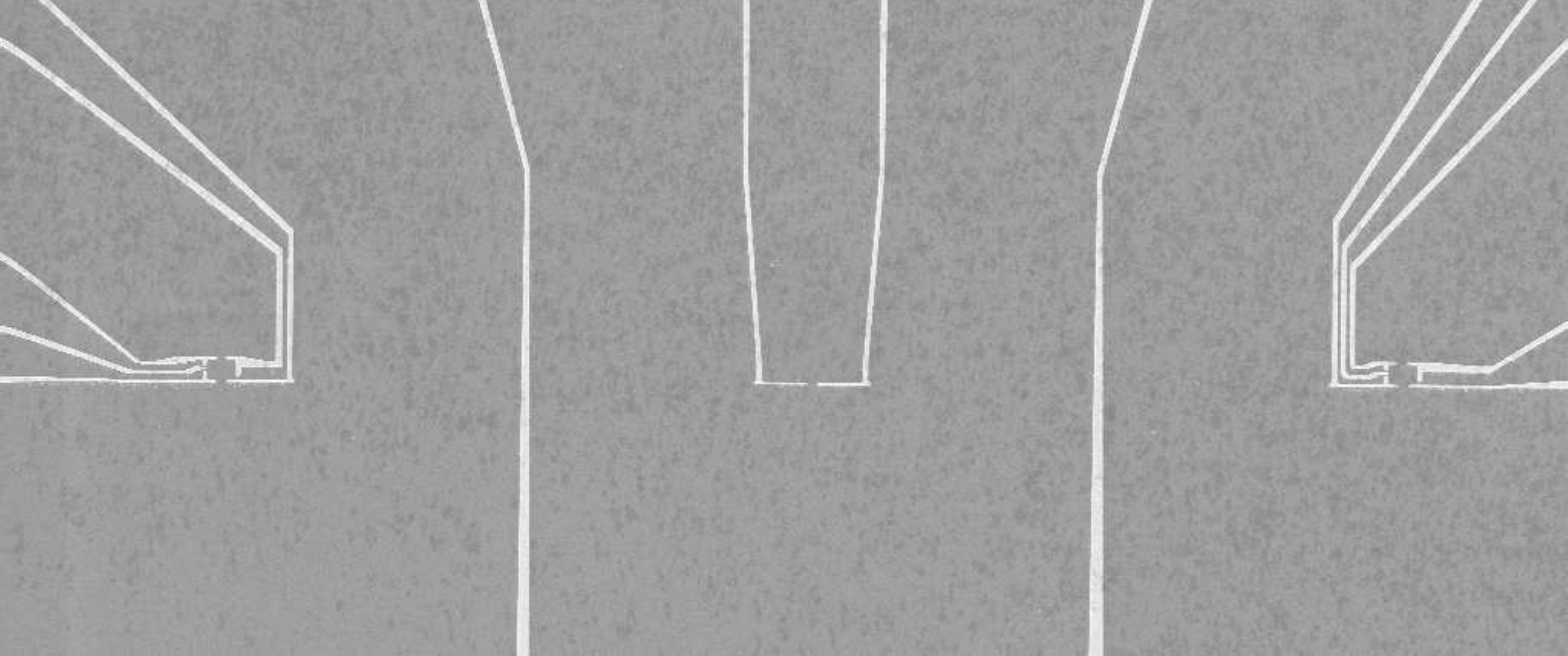}
\par\end{centering}

\caption{Device designed to explicitly test for heat leakage across a depleted
barrier. When the vertical gates are energized, the 2D systems on
the left and right are completely isolated (electrically) from the
2D system with the heater in the center.}
\end{figure}
\begin{figure}[H]
\begin{centering}
\includegraphics[scale=0.5]{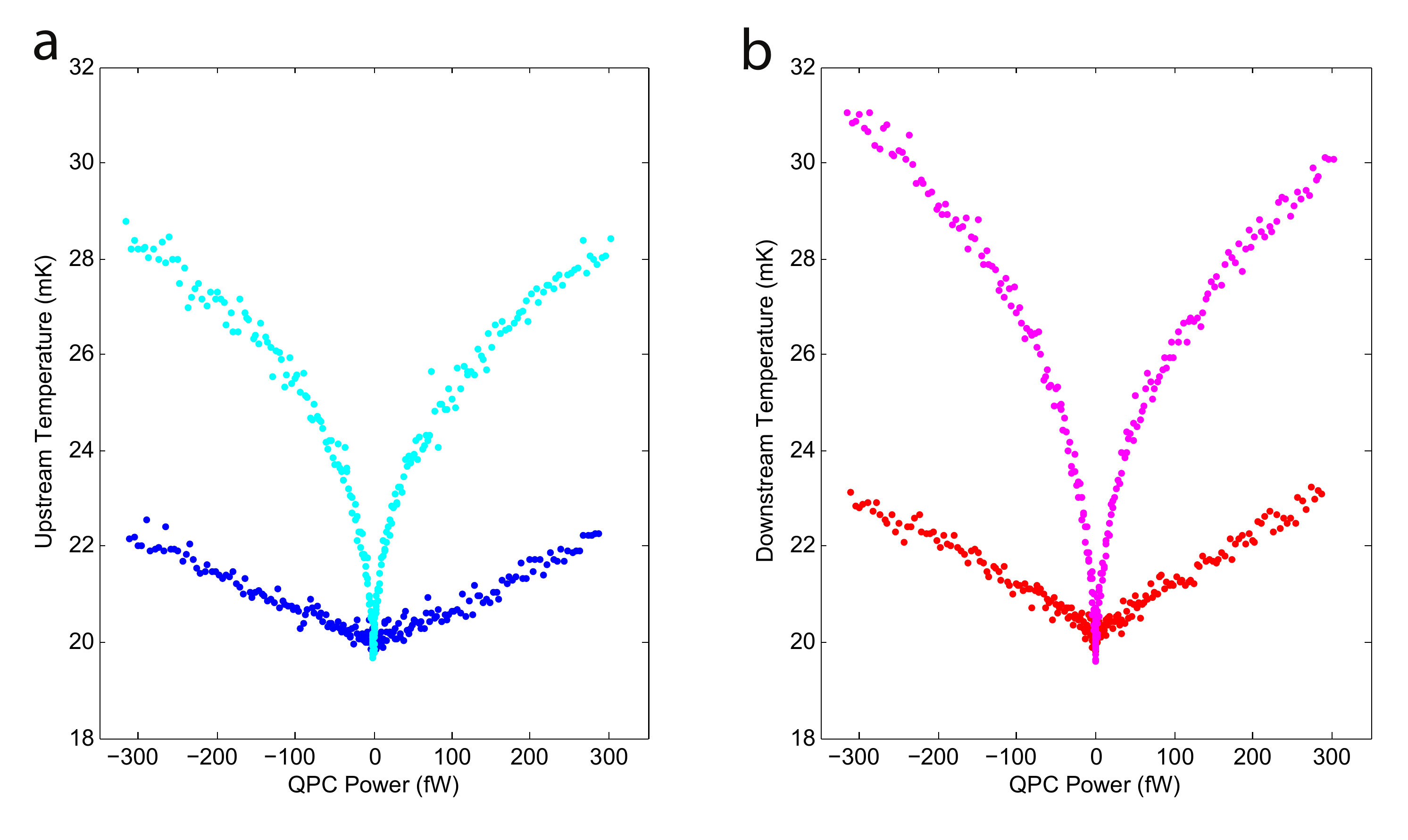}
\end{centering}

\caption{Heat transport across a depleted barrier. The cyan curve depicts temperature
measured upstream from the heater when the vertical gates are deenergized.
The blue curve depicts the temperature when the vertical gate is energized,
so heat must diffuse across a depleted region. The magenta and red
curves are the corresponding traces for the downstream dot. All data
was taken at 8.3 T, corresponding to column IV in Figure 3 of the
main paper.}
\end{figure}

\subsection*{\medskip{}
II. Geometric Diffusion Considerations}

Because we only detect neutral mode heating when there is a bulk contribution
to the heating signal, we have to ensure that there is no significant
change in the bulk contribution as we energize and deenergize the
deflector gate. It would appear plausible, for instance, that by turning
on the deflector gate we reduce the area over which the bulk heat
can diffuse. Specifically, with the deflector gate on, heat can no
longer diffuse up into the 2D region between our heater and our thermometers.
As a result, one may conjecture that more heat will be directed towards
the thermometers resulting in a higher temperature unassociated with
quantum Hall edge physics. The first indication that this redirection
of heat isn't relevant is the above observation (from CB peak shapes
and direct measurements) that heat does indeed partially diffuse through
depleted regions. A more convincing test, however, consists of altering
the geometry of the bulk to reduce the effect of this geometric distortion.

To this end, consider the device shown in Figure S11b. It is identical
to the devices used for measurements in the main body of the paper,
but with a shorter deflector gate length (8 um instead of 15 um).
At 8.3 T, we expect an edge structure as shown in Figure S11, with
two separated edges: one corresponding to the boundary between vacuum
and $\nu=\frac{2}{3}$ and the other corresponding to the boundary
between $\nu=\frac{2}{3}$ and $\nu=1$, as $\nu=1$ is the bulk filling
factor and $\nu=\frac{2}{3}$ is the edge that we detect with our
local injection measurements. In the 8 um deflector device, with the
deflector deenergized, we measure a slightly elevated resistance ($1.19\, R_{K}$),
indicating that the inner edge corresponding to the transition from
$\nu=\frac{2}{3}$ to $\nu=1$ is being backscattered (transmission
coefficient of $52\%$ for that inner edge). This indicates that the
$\nu=1$ bulk is largely closed off in this deflected region, so we
would expect very little bulk heat to diffuse upwards through this
narrow constriction. If the difference in upstream heating displayed
in Figure 3(IV) of the main paper is due to a redirection of bulk
heat flow, we would expect almost the same difference between the
temperature measured in the 15 um deflector device (Figure S11a) and
the 8 um deflector device (Figure S11b). 

The data from these measurements are presented in Figure 12. The blue
and red points correspond to temperatures measured in the device from
Figure S11a with deflectors off. The cyan and magenta points correspond
to temperatures measured in the device from Figure S11b, also with
deflectors off. These undeflected temperatures in the two devices
are very close, to within the data spread. For reference, the temperature
associated with turning on the deflectors (which results in the same
geometry for the two devices) is displayed in green and orange.

From these, we can infer that the excess temperature found in the
green and orange traces is indeed associated with a hot $\nu=\frac{2}{3}$
edge, as this edge is the only component of the system that is significantly
altered as deflector gates are turned on in the device from Figure
S11b.

In SOM D, we present yet another device, where the gate-defined edge
is replaced by a sharp mesa-defined edge. If the excess upstream heat
was due to a redirection of bulk heating, we would expect an elevated
temperature in that situation, given that the device possesses a nearly
identical bulk geometry to the gate-defined system. Here, however,
we don't see any heat associated with the edge at 6.2 T (Fig. S14a).
This provides even further evidence that the observed upstream heat
is due to FQH edge structure and is independent of the measured bulk
heat transport at $\nu=1$.

\begin{figure}[H]
\begin{centering}
\includegraphics[scale=0.25]{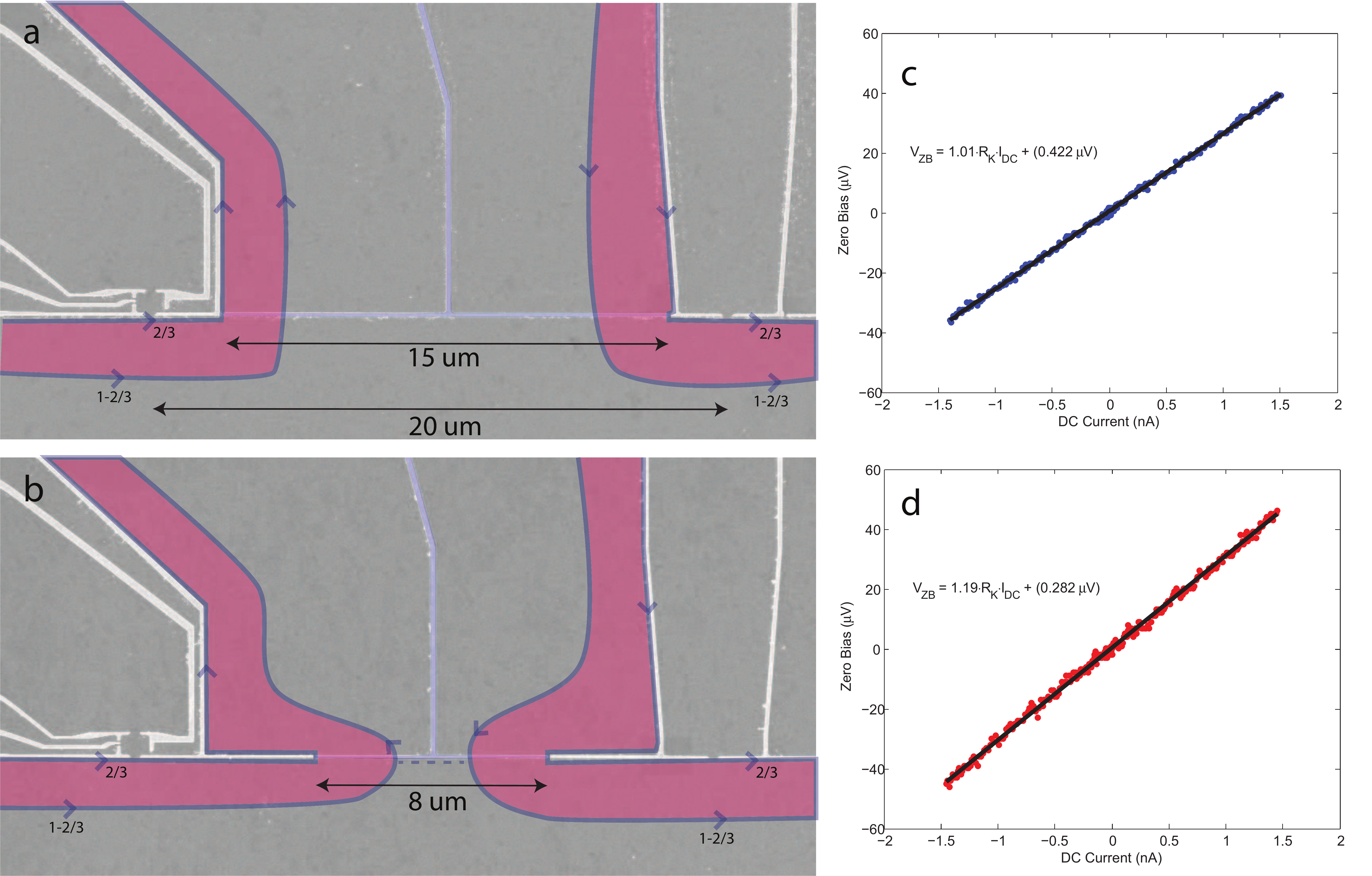}
\par\end{centering}

\caption{A device to test the effect of geometric diffusion considerations.
\textbf{a)} SEM image of device identical to that used for data in
the main paper. Edge labels correspond to what is expected at a field
of 8.3T, based on our local and global $R_{xy}$ measurements. \textbf{b)
}SEM image of a device with a narrower region through which edges
can be deflected. From the elevated resistance shown in panel d, we
know that the inner edge is partially backscattered. \textbf{c) }Copy
of the $\Delta V_{ZB}$ versus $I_{DC}$ curve from Figure S1, demonstrating
that the resistance in the deflector channel is the same as the bulk
value ($1.01R_{K}$), indicating that the $\nu=1$ state is fully
connected from the top to the bottom of the image in panel a. \textbf{d)
}A corresponding $\Delta V_{ZB}$ versus $I_{DC}$ curve for the device
in panel b. The elevated resistance ($1.19R_{K}$) indicates that
the inner edge, which has a conductance of $\frac{e^{2}}{3h}$ is
52\% transmitted. This suggests that the $\nu=1$ state is connected
through a narrow channel in this device, providing much less room
for heat to diffuse upwards compared to the device in panel a}
\end{figure}
\begin{figure}[H]
\begin{centering}
\includegraphics[scale=0.5]{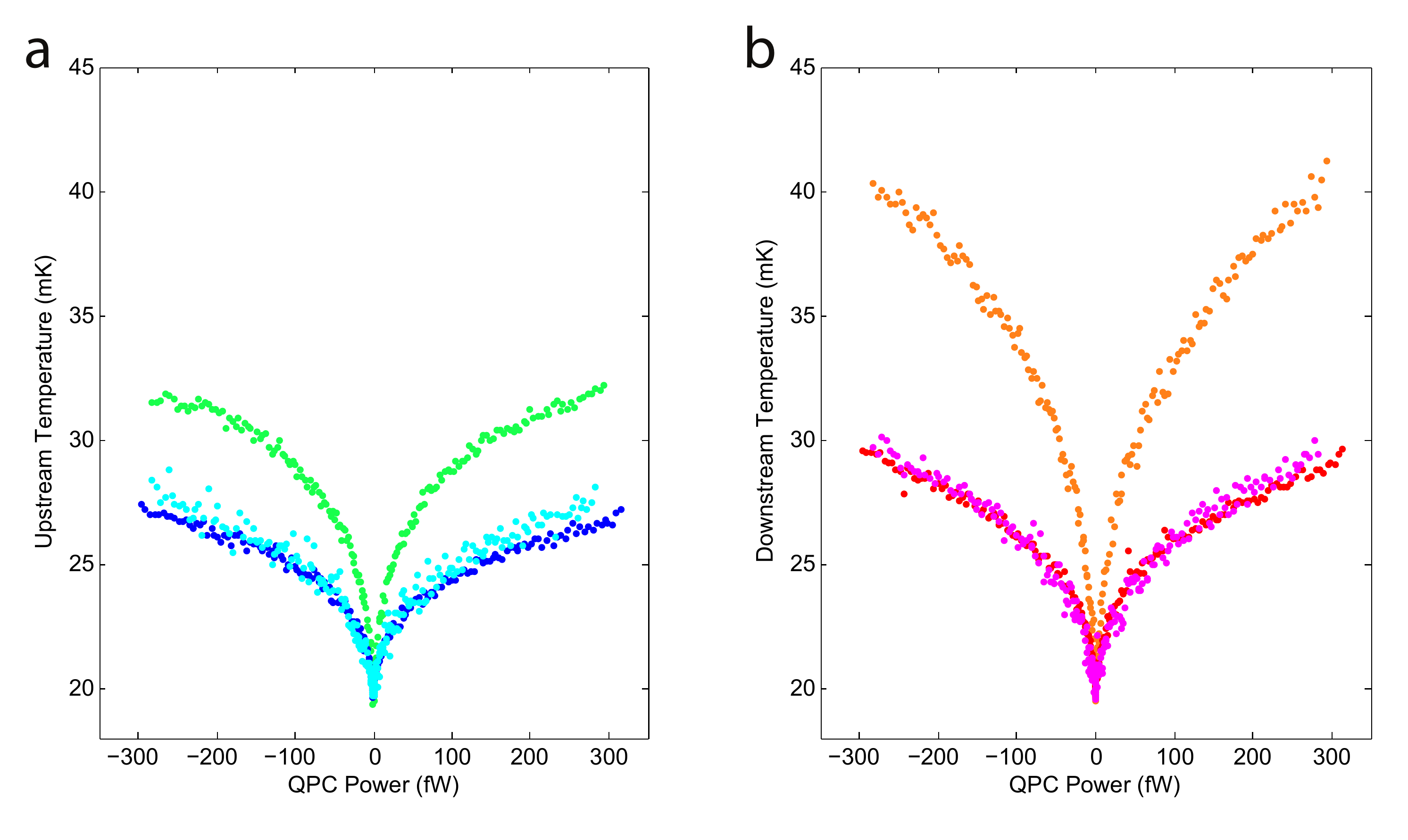}
\par\end{centering}

\caption{The blue and red points correspond to temperatures measured in the
device from Figure S11a with deflectors off. The cyan and magenta
points correspond to temperatures measured in the device from Figure
S11b, also with deflectors off. These undeflected temperatures in
the two devices are very close, to within the data spread. For reference,
the temperature associated with turning on the deflectors (which results
in the same geometry for the two devices) is displayed in green and
orange. From this we can conclude that the observed upstream heating
is not due to a redirection of bulk heating upon energizing of deflector
gates. \textbf{a) }Upstream. \textbf{b)} Downstream}
\end{figure}
\pagebreak{}

\section*{Supplement D: The Fractional Quantum Hall Edge at $\nu=1$}

The spatial separation between compressible edges is determined largely
by the sharpness of the confining potential. At $\nu=1$, the presence
of FQH edge structure requires a shallow confining potential (compared
to the magnetic length or Fermi wavelength), as well as a high mobility
2DES (as is always required for FQH physics). In this supplement,
we will present data from an edge defined through a mesa etch, creating
a steeper confining potential than what was presented in the main
paper. As a result of the steeper confinement, we end up with edges
of type III and IV (from Fig. 4) when the bulk is at $\nu=1$. The
gate-defined edge, as a reminder, had edges of type IV and V at bulk
filling $\nu=1$. From the table in Fig. 4, we can see that charge
transport ($R_{xy}$ and $R_{L}$) cannot discriminate between the
type III and type IV edges. In this supplement, we will present evidence
that both types of edge can exist in a single sample, and that they
can be distinguished by monitoring upstream heat transport.

\begin{figure}[H]
\begin{center}
\includegraphics[scale=0.22]{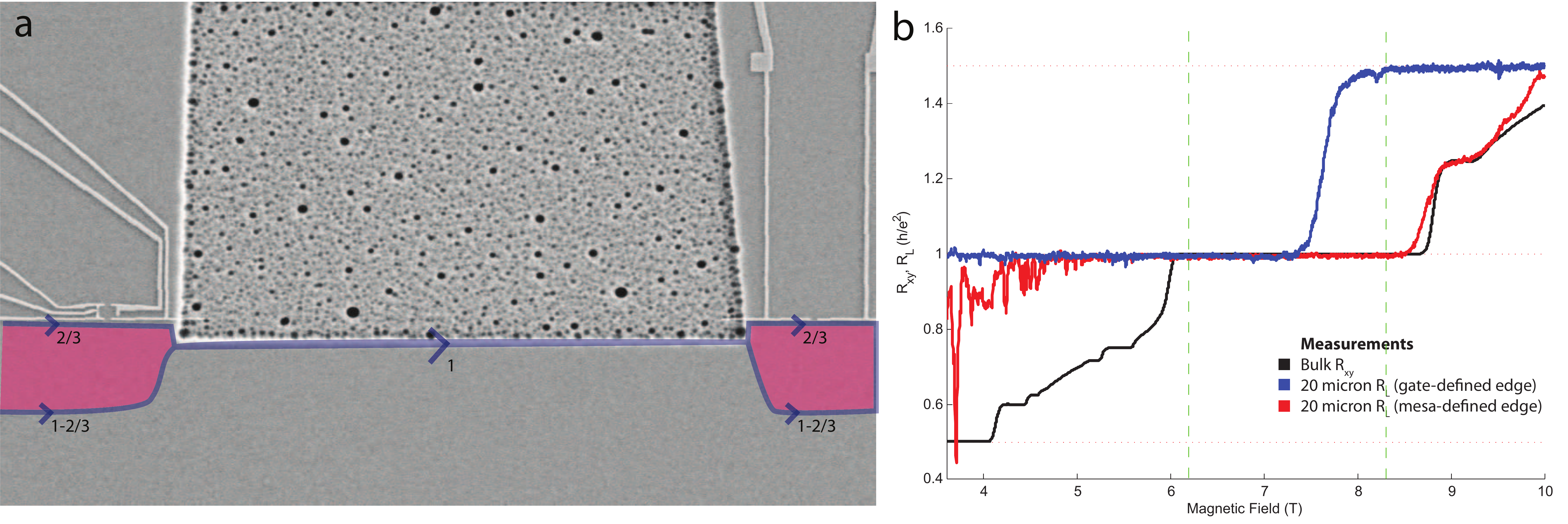}
\caption{\textbf{a) }Modified device to study heat transport along a sharper
edge. The gate defined edge (studied in the main paper) allowed for
a $\nu=2/3$ edge to form outside the $\nu=1$ bulk (blue trace in
panel \textbf{b}). The mesa-defined edge here is sharper, and the
sharp density gradient may precludes FQH edge structure outside the
$\nu=1$ bulk. This image is of a device with 40 $\mu$m between heater
and thermometer, while the device measured had 20 $\mu$m between
heater and thermometer, to match the device presented in the main
paper. \textbf{b)} $R_{L}$ (red) for the device in panel \textbf{a}.
The reduced resistance of the edge (red versus blue) at 8.3 T when
switching from a gate-defined to a mesa-defined edge suggests that
the originally separated FQH ($\nu=2/3$ and $\nu=1\rightarrow2/3$)
channels are brought close together, allowing charge to equilibrate
between them. While the device is drawn with an edge of type III (from
Fig. 4 of the main paper), an edge of type IV cannot be ruled out
from charge transport, either locally ($R_{L}$) or globally ($R_{xy}$).}
\end{center}
\end{figure}

In Figure S13, we present an SEM image of the device under consideration.
The device geometry and substrate used are identical to those used
for the device presented in the main paper. Using a wet-etching procedure,
we are able to remove material between the QPC heater and the QD thermometer.
This creates a physical boundary to the sample along which the edge
propagates. The density in the 2DES must drop to zero across this
edge, which can be happen over a shorter length scale than for an
edge created by depleting the 2DES via electrostatic gating. 

To demonstrate that this edge is sharper, we can repeat our local
charge transport measurements (Fig. S13b, $R_{L}$ in red). The observed
enhanced conductance at any given field (red compared to blue) is
a result of either more edges participating in transport, or a greater
conductance of those edges participating. This is precisely what is
expected if the edges are confined with a steeper potential. Here
we will focus on behavior on the edge of the $\nu=1$ bulk (6.2 T
and 8.3 T). From the charge transport measurements, we cannot distinguish
the exact edge structure at either field (see edges III and IV in
Fig. 4). 
\begin{figure}[H]
\begin{center}
\includegraphics[scale=0.23]{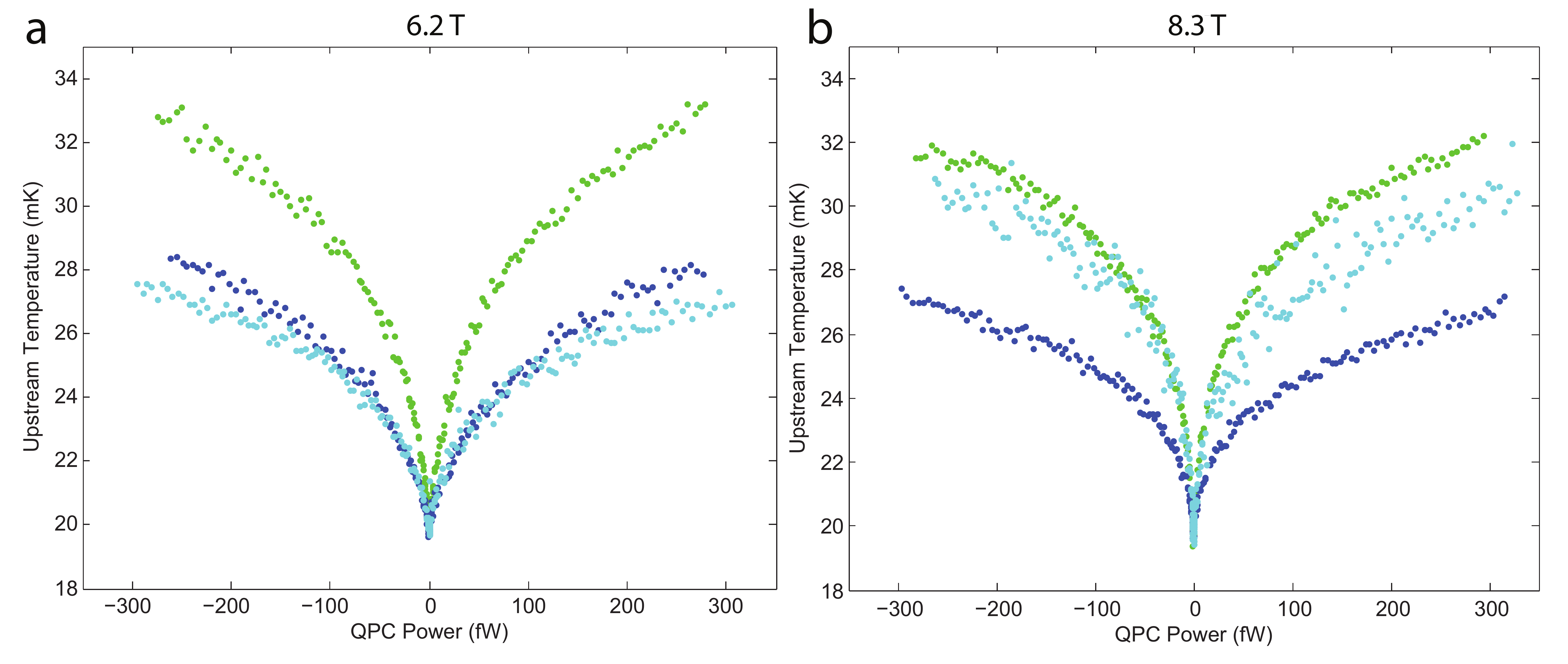}
\caption{Upstream thermometry to identify FQH structure in the $\nu=1$ edge.
Dark blue curves depict background upstream heating, which we attribute
to the bulk. The green curve depicts the heat observed with a gate-defined
edge connecting heater and thermometer. The light blue curve depicts
upstream heat observed with a sharper mesa-defined edge connecting
heater and thermometer. \textbf{a) }At low fields, the upstream heating
from the mesa-defined edge closely matches the background, suggesting
no excess heat is carried by the edge. \textbf{b)} At high fields,
there is a similar amount of upstream heating by both sharp and shallow
edges, both appreciably above the background.}
\end{center}
\end{figure}

To distinguish between these two possible edge structures, we can
perform upstream thermometry measurements. Because we have created
our edge via etching the mesa, we cannot control for bulk heating
by energizing and deenergizing deflection gates. However, by using
an identical geometry to the gate-defined device, we can still identify
the presence or absence of excess heat due to the edge. This thermometry
measurement is presented in Fig. S14, with data from the edge-defined
device in light blue. For comparison, data from the gate-defined device
taken at the same fields is reproduced in dark blue and green (identical
to upstream data in columns III and IV of Fig. 3 in the main paper).

At 6.2 T, we see that the temperature detected upstream (light blue)
closely matches the temperature associated with bulk heating in the
original device (dark blue). This is consistent with no heat being
transported by the edge. The lack of upstream heat carried by the
edge allows us to classify it as a simple IQH $\nu=1$ edge (type
III in Fig. 4), similar to what was observed at bulk fillings of $\nu=2$
and $\nu=3$ in the original device. 

At 8.3 T, the temperature measured upstream (light blue in Fig. S14b)
appears to be elevated, closely matching the temperature seen when
a $\nu=2/3$ edge connects the heater to the upstream thermometer
in the original device (green curve). Recall that in the original
device, this $\nu=2/3$ edge was detectable via measurement of $R_{L}$
(blue curve in Fig. 2). Here the charge signature has vanished ($R_{L}=R_{xy}$),
but the nearly identical upstream heating strongly suggests that the
$\nu=2/3$ edge is still present (edge IV in Fig. 4). These measurements
increase our confidence in assigning edge IV to our observations at
6.2 T in the original device.

\end{singlespace}
\end{document}